\RequirePackage{fix-cm}
\documentclass[smallextended]{svjour3}       
\smartqed  
\usepackage{graphicx}
\usepackage{framed}
\usepackage{natbib}
\usepackage{booktabs}
\usepackage{multirow} 
\usepackage{listings}
\usepackage{subcaption}
\usepackage{float}
\usepackage{rotating}
\usepackage{verbatim}
\usepackage{url}
\usepackage[most]{tcolorbox}
\begin{document}

\title{Do Code LLMs Do Static Analysis?
}


\author{Chia-Yi Su         \and
        Collin McMillan 
}


\institute{Chia-Yi Su \at
              Department of Computer Science and Engineering, University of Notre Dame, IN, USA. \\
              \email{csu3@nd.edu}           
           \and
          Collin McMillan \at
              Department of Computer Science and Engineering, University of Notre Dame, IN, USA.
}

\date{Accepted at Empirical Software Engineering}

\maketitle

\begin{abstract}
This paper investigates code LLMs' capability of static analysis during code intelligence tasks such as code summarization and generation.  Code LLMs are now household names for their abilities to do some programming tasks that have heretofore required people.  The process that people follow to do programming tasks has long been understood to require static analysis.  For example, human programmers navigate the call graph of large programs to comprehend the different parts of those programs.  Education in programming includes static analysis under the assumption that better static analysis skills beget better programming.  While popular culture is replete with anthropomorphic references such as LLM ``reasoning'', in fact code LLMs could exhibit a wholly alien thought process to humans.  This paper studies the specific question of static analysis by code LLMs.  We use three different static analysis tasks (callgraph generation, AST generation, and dataflow generation) and three different code intelligence tasks (code generation, summarization, and translation) with two different open-source models (Gemini and GPT-4o) and closed-source models (CodeLlaMA and Jam) as our experiments.  We found that LLMs show poor performance on static analysis tasks and that pretraining on the static analysis tasks does not generalize to better performance on the code intelligence tasks and vice versa.
\keywords{static analysis \and code intelligence tasks \and large language models\and LLM reasoning}
\end{abstract}

\section{Introduction}

Popular discussion around large language models (LLMs) is replete with anthropomorphic references such as LLM ``reasoning'' and ``creativity''~\citep{tian2024thinking,zhao2024assessing}.  Even dialogue in scientific literature uses biological terms such as ``attention'' and ``hallucination'' to refer to parts of LLM architecture or behavior.  The implication is that LLMs think and act like people, with human intentions and human thought processes.  One may conclude that this implication is clearly hyperbole -- many authors argue that impressive LLM abilities are in fact merely pattern matching across big data input~\citep{wang2023can} and/or carefully curated finetuning~\citep{chen2024automated}.  Yet the processes that LLMs follow are controversial and there is not yet sufficient empirical evidence on specific LLM behaviors to draw strong conclusions.

One group of LLM behaviors in the spotlight is in software development tasks.  Code LLMs are language models specializing in these tasks such as generating code samples from natural language prompts, code comment generation, and translation from one programming language to another.  Many programmers use code LLMs every day as a normal part of their workflow.  For example, Google estimated in November of 2024 that over 25\% of its new code is written by AI and later reviewed by engineers~\citep{arstechnica2024google}.  The rapid rise of LLM-generated code has led to anxiety about the safety and security of programs containing that code~\citep{mohsin2024can,wang2024your}, privacy and other legal issues~\citep{he2024emerged}, and even the future of software engineering as a career~\citep{rasnayaka2024empirical, tona2024exploring}.  Knowing how code LLMs do coding tasks could shed light on these concerns.

Meanwhile, a fundamental part of what people do during programming tasks is static analysis.  Static analysis is the process of examining code without executing it~\citep{ayewah2008using}.  Studies of human programmers have for decades shown how people build a mental model of program behavior through static tracing of three relationships in particular: function invocations (i.e., function call relationships), data flow among variables, and the hierarchical structure of code elements, such as how expressions, statements, and blocks are organized~\citep{littman1987mental}.  A rich body of literature describes these and similar static analysis tasks which underpin development of static analysis tools built into popular IDEs~\citep{luo2021ide}.  

This paper asks the question: do code LLMs do static analysis?  Humans do static analysis as a core part of software development, so if code LLMs follow a process similar to humans, then one would expect code LLMs to do static analysis.  We conduct experiments in which we 1) observe the performance of LLMs on software development tasks, 2) observe the performance of those same LLMs on static analysis tasks, 3) finetune open-source LLMs to do static analysis tasks, 4) observe if the finetuned LLMs' performance on development tasks changes, and 5) observe if LLMs have the byproduct as the results of being trained on code intelligence tasks.  One would expect improved performance on development tasks if the code LLMs are trained to do static analysis better and have static analysis as a byproduct.

We found that code LLMs generally do a poor job at static analysis tasks but have strong performance at coding tasks.  We also found that finetuning LLMs with static analysis tasks improved models' ability to do simple static analysis tasks, but did not improve models' ability on more complicated static analysis and coding tasks. In addition, training LLMs on code intelligence tasks does not lead to better generation of static analysis tasks.  We found these results over two open-source (CodeLlaMA, jam) and two closed-source models (GPT-4o, Gemini) and two programming languages (Java, C/C++).  We studied three static analysis tasks (call graph, data flow graph, and abstract syntax tree generation) and three programming tasks (code summarization, code translation, and code generation).  Our findings contribute to the ongoing academic debate about how LLMs function and have implications for task generalization, training procedures, and the design of future LLMs for software engineering tasks.


\section{Background and Related Work}

This section discusses studies of program comprehension and the role static analysis plays in that comprehension process, code large language models, and studies of capabilities of those models.

\subsection{Static Analysis in Program Comprehension}\label{sec:staticanalysis}

This section discusses the studies of program comprehension and static analysis for program comprehension.

\textbf{Empirical Studies} Program comprehension has been a target of empirical studies for a decade. Several empirical studies have been conducted to understand how programmers comprehend code.~\cite{littman1987mental} found that programmers  use data flow for comprehending code.~\cite{pennington1987stimulus} also found that programmers combine data flow and control flow to comprehend source code in their study with 80 professional programmers.~\cite{shneiderman1979syntactic} have similar conclusion that syntactic information is as important as semantic information. The way that programmers represent the code mentally can be referred as mental models~\citep{siegmund2016program, o2003software, storey2005theories, harth2017program}.~\cite{harth2017program} classified this into three different periods: 1) classical period 2) optimistic period 3) pragmatic period. In the classical period, several different strategies on how programmers understand code were proposed. During the optimistic period, several tools and studies were developed and conducted. The pragmatic period focus more on measurement instead of creating new methods.

\textbf{Static Analysis} Various studies have established that three static analysis tasks in particular are crucial to program comprehension: 1) call graph analysis, 2) data flow analysis, and 3) Abstract Syntax Tree (AST) generation. For example,~\citet{jiang2017programmers} conducted extensive studies on impact analysis tasks. They recruited nine professional programmers for a depth study and 35 programmers for the breath study. They found that callgraphs help programmers to navigate the source code.~\cite{wallace2025programmer} conducted a project-level eye-tracking study with 10 Java programmers for source code summarization. They suggested using callgraph to develop context-aware LLMs for source code summarization.~\cite{konopka2018data} conducted an empirical study with novice and intermediate programmers and showed the importance of using data flow analysis for program comprehension tasks. In addition, they proposed a data-flow based metrics to evaluate how human programmers comprehend a program.~\cite{bergeretti1985information} also show that the information flow helps various stages of software development.~\cite{maalej2014comprehension} conducted a study on how software developers comprehend source code with 28 software developers. They found that some participants drew the data flow to comprehend the program.~\cite{hovemeyer2016control} showed that simple AST helps to understand the strategies that students use for their programming assignments.~\cite{agrahari2020astar} show that AST helps the novice programmers to learn the fundamental of data structure.~\cite{schanzer2019accessible} developed an AST-based tool for vision impaired programmers. They found the improvement with their tools. Overall, these show that programmers rely on static analysis to do program comprehension tasks.


Static analysis helps programmers analyze code base without the need to execute the source code and any input. Different theories on static analysis have been proposed.~\cite{park2021survey} proposed the parametric static analysis to theorize the static analysis. Static analysis helps programmers in different aspects of software development~\citep{vassallo2020developers, beller2016analyzing}.~\cite{vassallo2018context} found that developers use the warning message from automatic static analysis tools to find the bugs in the source code.~\cite{ayewah2008using} combined the static analysis tools for bug detection. Static analysis can be used to aid the program comprehension~\citep{eisenbarth2001aiding}. In the study conducted by~\cite{tymchuk2018feedback}, they showed that static analysis tools can also be served as an educational tool.~\cite{singh2017evaluating} showed that static analysis tools can reduce the workload of code review.

\vspace{-2mm}
\subsection{Code Large Language Models}

Code LLMs have been the trend of in the recent year~\citep{zhang2023survey, jiang2024survey, chen2025an} and have been applied to several different software engineering tasks~\citep{nam2024using, gu2023llm, zheng2025towards, he2025llmbased, zhong2024can}. In this section, we discuss the history of code LLMs. Specifically, we focus on how LLMs are trained and how the source code is represented during different stages. Then, we discuss the application of LLMs in software engineering and studies on the capabilities of code LLMs.

\textbf{History} \cite{sun2024survey} divided the history of code LLMs into three groups: 1) neural language models for code 2) code pre-trained models 3) code LLMs. Natural language models represent the code as the natural language. Different methods have been proposed to represent code.~\cite{alon2019code2vec} encode each code token as a word vector.~\cite{Zhang2019Novel} showed that ASTs can better encode the information in the code.~\cite{leclair2020improved} applied graph neural network to represent code for code summarization. Code pre-trained models pretrain the language models with huge amount of source code and finetuned it with downstream tasks.~\cite{wang2021codet5} pretrained the models with encoder-decoder transformer architectures.~\cite{li2023starcoder} proposed a decoder architecture only model for pretrainig. More recently, the attention shift toward scaling up the models and in-context learning. The models include open source models Codellama~\citep{roziere2023code} and several closed source models~\citep{brown2020language, anil2023palm}.

\textbf{Applications} LLMs have been applied to different domains of software engineering tasks~\citep{zheng2025towards, zheng2023survey, jin2024llms}.~\cite{sarsa2022automatic} and~\cite{sun2023automatic} examine the capabilities of LLMs on code summarization.~\cite{khan2022automatic} applied LLMs to generate the summarization.~\cite{Koziolek2024llmbased} leveraged the LLMs and retrieval augmentation for code generations. Different evaluation methods have also been studied~\citep{liu2024your, liu2024exploring}.~\cite{yuan2024transagent} proposed  multi-agent systems for code translation.~\cite{pan2023stelocoder} applied decoder-only LLMs to translate different programming languages into Python. In addition to program comprehension tasks,~\cite{Xue2024llm4fin} used LLMs to generate the test case for FinTech software testing.~\cite{lu2024grace} use LLMs for vulnerabilities detection with in-context learning.~\cite{yin2024thinkrepair} and a LLM-based framework for automatic program repair. Overall, LLMs are widely used in the software engineering.

\textbf{Capability Studies} Several studies on LLMs capability have been published~\citep{yang2024if, li2024personal, jiang2024survey, fan2024exploring}.~\cite{yang2024if} divided the capabilities of code LLMs into three different ares 1) programming skills 2) complex reasoning 3) structure knowledge.~\cite{li2022competition} showed that LLMs defeat around half of the programmers during programming competitions. Several studies have showed the promising results on using multiple LLMs as an agent for code generation~\citep{liu2023dynamic, liu2310dynamic,talebirad2023multi}.~\cite{li2023hitchhiker} showed that LLMs can do complex reasoning on debugging.~\cite{chen2022program} showed that LLMs can reason the output of the code for several math problems. Although several papers showed promising results on reasoning, several studies have also showed that LLMs can only do reasoning when the code is simple~\citep{liu2024codemind, chen2024evaluating}.~\cite{madaan2022language} and~\cite{wang2022code4struct}  showed the ability of LLMs for structure prediction after few-shot learning. 

\section{Experimental Design}

Our experiments center around four topics: 1) we observe LLM performance on code development tasks, 2) we observe LLM performance on static analysis tasks in baseline, in-context learning setting, and finetuned setting, 3) we finetune LLMs to do static analysis and determine if those models' performance on code development tasks increases, and 4) we observe whether LLMs can do static analysis as a byproduct of being trained on code development tasks. This section covers our research questions, methods to answer those questions, and experimental settings.

We divide our experiments across open-source and closed-source LLMs.  For open-source LLMs we can closely control experimental variables and conduct finetuning.  For closed-source LLMs we can access the largest available models, but we give up control over many variables and the ability to finetune.

\subsection{Research Questions}

Our research objective is to determine if LLMs that do code development tasks also ``know how'' to do static analysis and to determine if models' ability to do static analysis is connected to the models' ability to do code development tasks and vice versa.  We ask the following Research Questions (RQs):

\begin{description}
    \item[~~RQ1] What is the performance of closed-source LLMs on static analysis tasks as is?
    \item[~~RQ2] What is the performance of open-source LLMs on static analysis tasks when finetuned for those tasks?
    \item[~~RQ3] What is the performance of open-source LLMs on code development tasks when finetuned for those tasks, after they have been finetuned for static analysis?
    \item[~~RQ4] What is the performance of open-source LLMs on static analysis tasks after they have been finetuned for code intelligence tasks?
    \item[~~RQ5] What are the common error types in static analysis?
   
\end{description}


The rationale behind RQ1 is that state-of-the-art commercial LLMs are already known to perform code development tasks well~\citep{zheng2025towards}, but we do not yet know how well these models perform static analysis tasks.  The rationale behind RQ2 is that LLMs may be able to perform static analysis tasks if they are explicitly trained to do so, and the availability of open-source models provides us the opportunity to do this training.  The rationale behind RQ3 is to examine whether finetuning the models with static analysis tasks could improve the performance of code intelligence tasks. This is because human programmers use various static analysis tasks as an intermediate thought process for code intelligence tasks (Refer to Section~\ref{sec:staticanalysis}). If a model truly has the anthropomorphic characteristic, static analysis should provide the fundamental knowledge for LLMs to perform code intelligence tasks. The rationale for RQ4 is that models can internalize some aspects of static analysis after finetuning models on code intelligence tasks as a byproduct if models can do static analysis tasks. This research question allows us to have a more thorough evaluation on whether LLMs can do static analysis tasks. The rationale for RQ5 is that LLMs may fall short on specific areas. RQ5 provides the insights on where the LLMs fall short on static analysis tasks. 

\subsection{Code Development Tasks}
\label{sec:devtasks}

In this section, we discuss the three different code development tasks that we use i.e. code summarization, code translation, and code generation and the datasets for each task. We follow the instructions from the dataset because our idea is to show whether LLMs follow what human programmers do for code development tasks. We use C/C++ and Java as an example.

\textbf{Code Summarization} Code summarization is the task of writing natural language descriptions of a section of code~\citep{haiduc2010on}.  Code summarization is important because it is the core of automatic documentation generation and related tool support for helping programmers understand code; the history of research efforts in code summarization has been chronicled in recent surveys~\citep{zhang2024review, zhang2022survey}.  We study this task in this paper because code summarization has become an advertised feature of code LLMs~\citep{sun2024source} and because different studies show how people do static analysis in order to do code summarization~\citep{rodeghero2015empirical, stapleton2020human}.  In this paper, for Java we use the ``170k'' dataset by~\citet{su2024distilled}.  That dataset has its origins in a study of code summarization dataset recommendations by~\citet{leclair2019recommendations} and further vetted by~\citep{bansal2021project}.  A major advantage to the dataset is that it's test set corresponds to a set held out of training in the \texttt{jam} model, which strongly avoids data contamination concerns (see Section~\ref{sec:codemodels}).  For C/C++ we use a dataset by~\citet{liu2021retrievalaugmented} which followed similar best practices (e.g.,~\citep{leclair2019recommendations, allamanis2019adverse}).

\textbf{Code Translation} Code translation is the task of writing a program in one language that has the same functionality as a program in another programming language. Code translation can help program comprehension by in converting pseudo code into machine readable code~\citep{holt1960man} and translate less-readable programming language into a more human-readable programming language~\citep{hong2024dont}. Several studies have shown that code translation requires static analysis when done manually by people~\citep{siddiqui2023towards, feautrier1991dataflow}.  In this paper, we use code translation dataset in CodeXGLUE~\citep{lu2021codexglue} as our dataset for Java. CodeXGLUE collects several datasets for code intelligence tasks. CodeXGLUE divides the datasets into code-code, code-text, text-code, and text-code. The code translation dataset that we use is collected from several public repositories with deduplication and clear train/test/validation separation. For C/C++, we use the C to Rust dataset in Huggingface website~\citep{ccodetrandataset}. For execution-based evaluation, we finetuned our models with CodeTransOcean~\citep{yan2023codetransocean} and test our models with datasets proposed by~\cite{pan2024lost} because we do not find test cases in CodeTransOcean at the time we download the dataset.

\textbf{Code Generation} Code generation is the task of generating source code that implements the functionality described in natural language. Code generation can be used as an educational tool to help novice programmers learn programming~\citep{becker2023programming,mosterman2006automatic} and is widely used to improve the productivity of software developers~\citep{ross2023programmer, xu2022inide}. Static analysis is part of generating code in that an understanding of the static structure leads to higher code quality and few bugs~\citep{barke2023grounded, nachtigall2019explaining}. In this paper, we used CONCODE dataset~\citep{iyer2018mapping} from CodeXGLUE~\citep{lu2021codexglue} for Java. The dataset is collected from 33,000 repositories and they filtered out the methods without Javadoc as natural language description. Although the dataset has very clear train/validation/test split, we use separate half of the validation set as our testset because we do not find the ground-truth for testset at the time we download. For C/C++, we use Xlcost~\citep{zhu2022xlcost}. Xlcost is a dataset that contains seven different programming languages and natural language, though we only use the C/C++ component. For execution-based evaluation, we finetuned our models with the programming competition dataset proposed by~\cite{li2022competition} and evaluated the results with pass@k and compilation rate. Note that the dataset includes the problems with several programming languages. We only used problems that have C/C++ solutions. We explain the metrics that we used in Section~\ref{sec:metrics}.  

\begin{table}[htbp]
    \centering
    \caption{Datasets and Models for Different Tasks in C/C++ and Java}
    \label{tab:datasets_models}

    \rotatebox{90}{%
        \begin{tabular}{c llccccc}
            \toprule
            \multirow{2}{*}{} & \multirow{2}{*}{Task} & \multirow{2}{*}{Language} &
            \multicolumn{2}{c}{Dataset} & \multicolumn{2}{c}{Models} \\
            \cmidrule(lr){4-5} \cmidrule(lr){6-7}
            & & & Name & Source & JAM & CodeLlaMA \\
            \midrule
            \multirow{9}{*}{\rotatebox{90}{Development~~~}}
            & \multirow{2}{*}{Code Summarization} & Java
            & 170k & \citet{su2024distilled} & x & x \\
            & & C/C++
            & \citet{liu2021retrievalaugmented} & \citet{liu2021retrievalaugmented} & ~ & x \\
            \cmidrule(lr){2-7}
            & \multirow{4}{*}{Code Translation} & Java
            & CodeXGLUE & \citet{lu2021codexglue} & x & x \\
            & & C/C++
            & C to Rust & \citet{ccodetrandataset} & ~ & x \\
            & & Java
            & Codetransocean & \citet{yan2023codetransocean} & ~ & x \\
            & & Java
            & PLTranslationEmpirical & \citet{pan2024lost} & ~ & x \\
            \cmidrule(lr){2-7}
            & \multirow{3}{*}{Code Generation} & Java
            & CONCODE & \citet{lu2021codexglue} & x & x \\
            & & C/C++
            & Xlcost & \citet{zhu2022xlcost} & ~ & x \\
            & & C/C++
            & Code Contest & \citet{li2022competition} & ~ & x \\
            \midrule
            \multirow{6}{*}{\rotatebox{90}{Static Analysis~~}}
            & \multirow{2}{*}{AST Generation} & Java
            & funcom & \citet{leclair2019recommendations} & x & x \\
            & & C/C++
            & \citet{haque2021action} & \citet{haque2021action} & ~ & x \\
            \cmidrule(lr){2-7}
            & \multirow{2}{*}{Call Graph Generation} & Java
            & funcom & \citet{leclair2019recommendations} & x & x \\
            & & C/C++
            & \citet{haque2021action} & \citet{haque2021action} & ~ & x \\
            \cmidrule(lr){2-7}
            & \multirow{2}{*}{Dataflow Graph Generation} & Java
            & funcom & \citet{leclair2019recommendations} & x & x \\
            & & C/C++
            & \citet{haque2021action} & \citet{haque2021action} & ~ & x \\
            \bottomrule
        \end{tabular}
    }
\end{table}

\subsection{Static Analysis Tasks}
\label{sec:statictasks}

In this section, we discuss three static analysis tasks in our experiments, i.e. AST, dataflow graph, and callgraph generation with two different programming languages, i.e. C/C++ and Java.

\textbf{Abstract Syntax Tree Generation} An Abstract Syntax Tree (AST) is a tree structure that captures the syntactic information of the source code. In addition to being a key intermediate representation in code compilation, ASTs are used to localize software bugs~\citep{neamtiu2005understanding}, comprehend source code~\citep{sun2023abstract,welty1997augmenting}, and in software evolution~\citep{neamtiu2005understanding}. Different studies have also shown that ASTs help program comprehension, such as when learning how to code~\citep{agrahari2020astar, egan2014program}. In this paper, for Java we use dataset recommendation by~\citet{leclair2019recommendations}. We use 52M to train \texttt{jam}~\citep{su2023language} and further extract 25k to train CodeLlaMA~\citep{roziere2023code} for practical reasons (refer to Section~\ref{sec:codemodels}). We use the same testset (i.e., the 8k testset) proposed by~\citet{leclair2019recommendations}. For C/C++, we extract 9k methods for test and 10k methods for training from a dataset proposed by~\citet{haque2021action}. We used this dataset because  they followed~\citet{leclair2019recommendations} for preprocessing. For both C/C++ and Java, we used the srcML tool~\citep{srcml} to generate AST in XML format in keeping with recommendations by~\citet{bansal2023function}.  

\textbf{Callgraph Generation} Callgraph is a graph that represents the relationship between each method~\citep{ryder1979constructing}. Callgraphs are an important information for code comprehension source code~\citep{alanazi2021facilitating, walunj2019graphevo, bansal2024programmer, alanazi2021software}. In this paper, we used Doxygen~\citep{doxygen} to generate callgraphs for both C/C++ and Java. For C/C++ dataset, we extracted 25k methods as our testset and 20k methods as our training set from~\citep{haque2021action}. We generated the callgraph within a file for C/C++ to reduce the tokens of the prompt. For Java, we used the testset proposed by~\citet{leclair2019recommendations} and further extracted 20k methods from the rest of dataset for finetuning. We generated the callgraph from the entire project because Java does not usually have the callgraph within the file. To reduce the tokens of the prompt, we only provided caller methods in the prompt instead of the complete files. 

\textbf{Dataflow Graph Generation} Dataflow graph represents how one variable passes to another method or variable. Dataflow graph is important because it helps programmers to do code comprehension~\citep{sihler2024improving,kargén2012inputtracer} and increase productivity of software developers~\citep{alsaiyd2017source}. In this paper, we used Joern~\citep{joern} to generate the data flow of the method. We used Joern because Joern is highly mature for Java and C/C++. For Java dataset, we used the testset recommended by~\citet{leclair2019recommendations} to avoid data contamination. We further extracted 15k methods other than testset as our training set. For C/C++, we extracted 8k methods as our testset and 8k methods as our training set from~\cite{haque2021action}. We used the same procedure to generate dataflow for both datasets.

\vspace{-2mm}
\subsection{Code Models}
\label{sec:codemodels}

We use two open-source models and two closed-source models in our experiments.  Our goal was to use a range of models from a small one where we can control all experimental variables, to large models that represent the state-of-the-art but where we lose control of several variables.

\vspace{2mm}
\hspace{-5mm}\emph{Open Source Models}
\vspace{1mm}

\textbf{JAM} We used Jam~\citep{su2023language} with 1,024 context length throughout out entire experiments. \texttt{jam} is a language model based on GPT2 architecture for various code intelligence tasks on Java. \texttt{jam} was pretrained with 52m Java methods with clear train/validation/test set, public available dataset, and deduplication toolkit to avoid data contamination. While \texttt{jam} only has 350m parameters,~\citet{su2024distilled} showed in their study that it is small enough to be useful for code intelligence tasks. The advantage of this model is that this model can be run on a single consumer hardware, e.g. 16G VRAM GPU and can be used for the experiments with more controls before scaling up to larger language models.

\textbf{CodeLlaMA Instruct} CodeLlaMA instruct~\citep{roziere2023code} is an open-souce model released by Meta. We used CodeLlaMA instruct because it can better understand the prompt. We used the 13B model in our experiments because it can better fit into our 24G GPU~\citep{weyssow2025exploring} when using Qlora~\citep{dettmers2024qlora}. We used the Qlora released in~\citep{qlora}. CodeLlaMA is the trade-off between closed-sourced models and Jam because we have access to the models although we still do not have the control on the dataset used for pretraining.   

\vspace{2mm}
\hspace{-5mm}\emph{Closed Source Models}
\vspace{1mm}

\textbf{GPT-4o mini} GPT-4o mini is the commercial models released by OpenAI~\citep{achiam2023gpt}. GPT-4o mini is a closed-source model namely we do not have access to the pretraining data and we send our private source code to the model through API. Because of its closed-source, we cannot avoid data contamination issues~\citep{balloccu2024leak, jiang2024investigating} and data privacy issues~\citep{wu2024unveiling}. Though several studies have shown the success of GPT models~\citep{jain2025assessing, wang2024enhancing}.

\textbf{Gemini} Gemini is another commercial model released by Google~\citep{team2023gemini, team2024gemini} and is enhanced for multimodal capability. Gemini is also a closed-source model. We used Gemini as another closed-souce model because Gemini has been compared with OpenAI's ChatGPT in several studies in both software engineering~\citep{sobo2025evaluating,qi2023gemini,siam2024programming, su2024context} and other research topics~\citep{strzalkowski2024evaluation} and has shown superior than GPT in some studies~\citep{lee2023gemini}.

\begin{table}[]
    \centering
    \small
    \caption{Configurations of fine-tuning and evaluation of different models for each RQ.  Note, jam is Java-only but for all other models we had separate Java and C fine-tuning and evaluation.}
    \label{tab:rqconfigs}
\begin{tabular}{lllll}
                                           &                            & \multicolumn{2}{c}{fine-tuning}               & evaluation      \\
\multicolumn{1}{l|}{}                      & model                      & static analysis       & code dev.             &                 \\ \hline
\multicolumn{1}{l|}{\multirow{5}{*}{RQ2}}  & \multirow{2}{*}{jam}       & ast gen.              & \multicolumn{1}{c}{-} & ast gen.        \\
\multicolumn{1}{l|}{}                      &                            & dataflow gen.         & \multicolumn{1}{c}{-} & dataflow gen.   \\ \cline{2-5} 
\multicolumn{1}{l|}{}                      & \multirow{3}{*}{codellama} & ast gen.              & \multicolumn{1}{c}{-} & ast gen.        \\
\multicolumn{1}{l|}{}                      &                            & call graph gen.       & \multicolumn{1}{c}{-} & call graph gen. \\
\multicolumn{1}{l|}{}                      &                            & dataflow gen.         & \multicolumn{1}{c}{-} & dataflow gen.   \\ \hline
\multicolumn{1}{l|}{\multirow{13}{*}{RQ3}} & \multirow{4}{*}{jam}       & ast gen.              & code sum.             & code sum.       \\
\multicolumn{1}{l|}{}                      &                            & dataflow gen.         & code sum.             & code sum.       \\
\multicolumn{1}{l|}{}                      &                            & ast gen.              & code gen.             & code gen.       \\
\multicolumn{1}{l|}{}                      &                            & dataflow gen.         & code gen.             & code gen.       \\ \cline{2-5} 
\multicolumn{1}{l|}{}                      & \multirow{9}{*}{codellama} & ast gen.              & code sum.             & code sum.       \\
\multicolumn{1}{l|}{}                      &                            & call graph gen.       & code sum.             & code sum.       \\
\multicolumn{1}{l|}{}                      &                            & dataflow gen.         & code sum.             & code sum.       \\
\multicolumn{1}{l|}{}                      &                            & ast gen.              & code trans.           & code trans.     \\
\multicolumn{1}{l|}{}                      &                            & call graph gen.       & code trans.           & code trans.     \\
\multicolumn{1}{l|}{}                      &                            & dataflow gen.         & code trans.           & code trans.     \\
\multicolumn{1}{l|}{}                      &                            & ast gen.              & code gen.             & code gen.       \\
\multicolumn{1}{l|}{}                      &                            & call graph gen.       & code gen.             & code gen.       \\
\multicolumn{1}{l|}{}                      &                            & dataflow gen.         & code gen.             & code gen.    \\ \hline

\multicolumn{1}{l|}{\multirow{9}{*}{RQ4}} 
                   
    & \multirow{9}{*}{codellama} 
    & \multicolumn{1}{c}{-}        & code sum.     & ast gen. \\

\multicolumn{1}{l|}{}                      
    &                                    
    & \multicolumn{1}{c}{-}  & code sum.     & callgraph gen. \\

\multicolumn{1}{l|}{}                      
    &                                    
    & \multicolumn{1}{c}{-}    & code sum.     & dataflow gen. \\

\multicolumn{1}{l|}{}                      
    &                                    
    & \multicolumn{1}{c}{-}         & code trans.   & ast gen. \\

\multicolumn{1}{l|}{}                      
    &                                    
    & \multicolumn{1}{c}{-}  & code trans.   & callgraph gen. \\

\multicolumn{1}{l|}{}                      
    &                                    
    & \multicolumn{1}{c}{-}    & code trans.   & dataflow gen. \\

\multicolumn{1}{l|}{}                      
    &                                    
    & \multicolumn{1}{c}{-}        & code gen.     & ast gen. \\

\multicolumn{1}{l|}{}                      
    &                                    
    &\multicolumn{1}{c}{-}  & code gen.     & callgraph gen. \\

\multicolumn{1}{l|}{}                      
    &                                    
    & \multicolumn{1}{c}{-}   & code gen.     & dataflow gen. \\

\end{tabular}

\end{table}

\subsection{Evaluation Metrics}
\label{sec:metrics}
In this section, we explain the evaluation metrics for three static analysis tasks and three code intelligence tasks.

\textbf{Levenshtein distance} Levenshtein distance calculates the similarity between two strings by computing the number of edits between two strings. In this paper, we used rapidfuzz implemented by~\citet{rapidfuzz} and converted the value to zero to one scale, where zero means two strings are dissimilar and one means two strings are the same.

\textbf{Pair Accuracy} Pair accuracy calculates the percentage of accurate edges in a predicted callgraph. For example, in a predicted callgraph with edges $E={(A,B),(B,C)}$ and the correct edge $E={(A,B)}$, the pair accuracy would be $0.5$ because there are two edges in the predicted callgraph and only one of the predicted edges is correct.

\textbf{Chain Accuracy} Chain accuracy calculates the percentage of accurate callgraphs in all predicted callgraphs. For example, in a predicted callgraph with edges $E={(A,B),(B,C)}$ and the true callgraph with edges $E={(A,B)}$, the chain accuracy would be $0$ because there are two edges in the predicted callgraph, but there is only one edge in the true callgraph.

\textbf{METEOR} METEOR~\citep{banerjee2005meteor} is a metric to evaluate summarization tasks that considers the word similarity rather than exact word match only. The scale of METEOR is between 0 and 1, where zero means no similarity between two sentences and one means two sentences are the same.

\textbf{USE} USE~\citep{haque2022semantic} encodes the reference summary and the predicted summary to USE score and computes the cosine similarity between reference summary and the predicted summary.~\citet{haque2022semantic} shows that USE align better with human perspective. The scale of USE score is between -1 and 1, where -1 means there is no similarity between two summaries and 1 means two summaries are most similar.

\textbf{CodeBERTScore} CodeBERTScore~\citep{zhou2023codebertscore} is based on the BERTScore. CodeBERTScore encodes both natural language description and source code into a fixed length vector and computes the cosine similarity between encoded representation of the reference and the predicted tokens. In this paper, we used the package provided by~\cite{zhou2023codebertscore} to compute the CoderBERTScore F1.

\textbf{Pass@k} pass@k~\citep{chen2021evaluating} evaluates how many generated samples passes test cases, where k is the number of generated samples. In this paper, we used $k=1$ as our evaluation. 

\textbf{Compilation rate} Unlike pass@k, compilation rate calculates how many generated samples are successfully compiled without the need to pass the test cases. The scale of the compilation rate is between 0 and 1, where zero means no generated code can be compiled and one means all of the generated code are compiled successfully.  

\vspace{-3mm}

\subsection{Prompt Templates}
\label{sec:prompttemplate}
 In this section, we explain the prompt template for static analysis tasks. We used the same prompts for both Gemini and GPT models. Please note that the main purpose of this paper is to show that current LLMs are lack of reasoning skills instead of finding the prompts with the highest performance.

\textbf{Template for dataflow analysis without examples} We used the following prompt for dataflow analysis without examples, where \{source\_code\} is the function for dataflow analysis. \{main\_
method\_name\} is the source of the dataflow and \{sink\} is the sink of the dataflow.

\begin{framed}
\noindent
{\fontsize{9}{10}\selectfont
\begin{lstlisting}[
basicstyle=\ttfamily\fontsize{8}{9}\selectfont,
breaklines=true,
breakatwhitespace=false,
breakindent=0pt,
breakautoindent=false,
columns=fullflexible,
keepspaces=true,
aboveskip=0pt,
belowskip=0pt,
xleftmargin=0pt,
framexleftmargin=0pt
]
Given the code {source_code}, please do the dataflow analysis from source {main_method_name} to sink {sink} and print the statements, but without any explanation and markdown. Also, treat each statement as a statment instead of a block. Return the result with the template result:<statement1> -> <statement2> result:<statement1> -> <statement2>. Remember to use template as well.
\end{lstlisting}
}
\end{framed}

\textbf{Template for dataflow analysis with example} We used the following prompt for dataflow analysis with the example, where \{source\_code\} is the function for dataflow analysis. \{main\_
method\_name\} is the source of the dataflow and \{sink\} is the sink of the dataflow. \{example\_code\} is the example function and \{example\_data\_flow\} is the dataflow of the example code.
\begin{framed}
\noindent
{\fontsize{9}{10}\selectfont
\begin{lstlisting}[
basicstyle=\ttfamily\fontsize{8}{9}\selectfont,
breaklines=true,
breakatwhitespace=false,
breakindent=0pt,
breakautoindent=false,
columns=fullflexible,
keepspaces=true,
aboveskip=0pt,
belowskip=0pt,
xleftmargin=0pt,
framexleftmargin=0pt
]
example_code = '''
int main(int argc, char *argv[]) {
    if (argc > 1 && strcmp(argv[1], "42") == 0) {
        fprintf(stderr, "It C!\n");
        exit(42);
    }
    printf("What is this programming language?\n");
    exit(0);
}    
example_data_flow = '''
result: int main(int argc, char *argv[]) { ->  
if (argc > 1 && strcmp(argv[1], "42") == 0) {
'''

Given the example code {example_code} and its dataflow analysis result {example_data_flow} from source main to sink strcmp. Given the code {source_code}, please do the dataflow analysis from source {main_method_name} to sink {sink} and print the statements, but without any explanation and markdown. Please follow the example to generate the dataflow analysis result.
\end{lstlisting}
}
\end{framed}

\textbf{Template for AST generation without example} We used the following prompt for AST generation without examples, where \{source\_code\} is the function for AST generation.
\begin{framed}
\noindent
{\fontsize{9}{10}\selectfont
\begin{lstlisting}[
basicstyle=\ttfamily\fontsize{8}{9}\selectfont,
breaklines=true,
breakatwhitespace=false,
breakindent=0pt,
breakautoindent=false,
columns=fullflexible,
keepspaces=true,
aboveskip=0pt,
belowskip=0pt,
xleftmargin=0pt,
framexleftmargin=0pt
]
Given the code {source_code}, please generate the complete srcml of the given code without any explanation and markdown. Please remember to follow srcml rules.
\end{lstlisting}
}
\end{framed}

\textbf{Template for AST generation with example} We used the following prompt for AST generation with examples, where \{source\_code\} is the function for AST generation. \{example\_code\} is the example function and \{example\} is the srcml of the example code.

\begin{framed}
\noindent
{\fontsize{9}{10}\selectfont
\begin{lstlisting}[
basicstyle=\ttfamily\fontsize{8}{9}\selectfont,
breaklines=true,
breakatwhitespace=false,
breakindent=0pt,
breakautoindent=false,
columns=fullflexible,
keepspaces=true,
aboveskip=0pt,
belowskip=0pt,
xleftmargin=0pt,
framexleftmargin=0pt
]
example = '''
<?xml version="1.0" encoding="UTF-8" standalone="yes"?><unit xmlns="http://www.srcML.org/srcML/src" revision="1.0.0" language="C" filename="test.c"><function><type><name>int</name></type> <name>main</name><parameter_list>()</parameter_list> <block>{<block_content><expr_stmt><expr><call><name>printf</name><argument_list> (<argument><expr><literal type="string">"Hello World!"</literal> </expr></argument>)</argument_list></call></expr>;</expr_stmt> <return>return <expr><literal type="number">0</literal></expr>;</return></block_content>}</block></function></unit>
'''
example_code = '''
int main() {
    printf("Hello World!");
    return 0;
}
'''

Here's the example code {example_code} and the srcml of that code {example}. Given the code {code}, please generate the complete srcml of that code. Please remember to follow the example. Here's the srcml:
\end{lstlisting}
}
\end{framed}

\textbf{Template for callgraph generation without example} We used the following prompt for callgraph generation without examples, where \{source\_co-de\} is the function for callgraph generation.
\begin{framed}
\noindent
{\fontsize{9}{10}\selectfont
\begin{lstlisting}[
basicstyle=\ttfamily\fontsize{8}{9}\selectfont,
breaklines=true,
breakatwhitespace=false,
breakindent=0pt,
breakautoindent=false,
columns=fullflexible,
keepspaces=true,
aboveskip=0pt,
belowskip=0pt,
xleftmargin=0pt,
framexleftmargin=0pt
]
Given the example code {example_code} and example callgraph {example_callgraph} for the function calculate. Given the code {source_code}, please follow the example to generate all paths of the callgraph for method {function_name} in the format path: a->b->c etc, where -> means call. please just generate the graph without any explanation, markdown, additional information and remember the format is path: a->b->c
\end{lstlisting}
}
\end{framed}

\textbf{Template for callgraph generation with example} We used the following prompt for callgraph generation with examples, where \{source\_code\} is the function for callgraph generation.~\{example\_code\} is the source code example for generating callgraph and the~\{example\_callgraph\} is the callgraph for the \{example\_code\}.

\newpage

\begin{framed}
\noindent
{\fontsize{9}{10}\selectfont
\begin{lstlisting}[
basicstyle=\ttfamily\fontsize{8}{9}\selectfont,
breaklines=true,
breakatwhitespace=false,
breakindent=0pt,
breakautoindent=false,
columns=fullflexible,
keepspaces=true,
aboveskip=0pt,
belowskip=0pt,
xleftmargin=0pt,
framexleftmargin=0pt
]
example_code = '''

int add(int a, int b) {
    return a + b;
}

int multiply(int a, int b) {
    return a * b;
}

int calculate(int x, int y) {
    int sum = add(x, y);
    int product = multiply(x, y);
    return sum + product;
}

int main() {
    int result = calculate(3, 4);
    return result;
}
example_callgraph = '''
    path: calculate -> add
    path: calculate -> multiply
'''
Given the example code {example_code} and example callgraph {example_callgraph} for the function calculate. Given the code {source_code}, please follow the example to generate all paths of the callgraph for method {function_name} in the format path: a->b->c etc, where -> means call. please just generate the graph without any explanation, markdown, additional information and remember the format is path: a->b->c
\end{lstlisting}
}
\end{framed}

\textbf{Template for AST generation after finetuning with code summarization} We used the following prompt for AST generation after finetuning with code summarization, where \{code\} is the function for AST generation. \{example\_code\} is the source code example for generating AST and the \{example\} is the AST for the \{example\_code\}.

\begin{framed}
\noindent
{\fontsize{9}{10}\selectfont
\begin{lstlisting}[
basicstyle=\ttfamily\fontsize{8}{9}\selectfont,
breaklines=true,
breakatwhitespace=false,
breakindent=0pt,
breakautoindent=false,
columns=fullflexible,
keepspaces=true,
aboveskip=0pt,
belowskip=0pt,
xleftmargin=0pt,
framexleftmargin=0pt
]
example = '''
<?xml version="1.0" encoding="UTF-8" standalone="yes"?><unit xmlns="http://www.srcML.org/srcML/src" revision="1.0.0" language="C" filename="test.c"><function><type><name>int</name> </type> <name>main</name><parameter_list>()</parameter_list><block>{<block_content><expr_stmt><expr><call><name>printf</name><argument_list>(<argument><expr><literal type="string">"Hello World!"</literal></expr></argument>)</argument_list></call></expr>;</expr_stmt><return>return <expr><literal type="number">0</literal></expr>;</return</block_content>}</block></function></unit>
'''
example_code = '''
int main() {
    printf("Hello World!");
    return 0;
}
'''
[INST] You have know how to describe the code. Now, here's the example code {example_code} and the srcml of that code {example}. can you follow the example to generate the srcml of the {code} without explanation -- just srcml [/INST]
\end{lstlisting}
}
\end{framed}

\textbf{Template for AST generation after finetuning with code generation} We used the following prompt for AST generation after finetuning with code generation, where \{code\}
is the function for AST generation. \{example\_code\} is the source code example for generating AST and the \{examp-
le\} is the AST for the \{example\_code\}.
\begin{framed}
\noindent
{\fontsize{9}{10}\selectfont
\begin{lstlisting}[
basicstyle=\ttfamily\fontsize{8}{9}\selectfont,
breaklines=true,
breakatwhitespace=false,
breakindent=0pt,
breakautoindent=false,
columns=fullflexible,
keepspaces=true,
aboveskip=0pt,
belowskip=0pt,
xleftmargin=0pt,
framexleftmargin=0pt
]
example = '''
<?xml version="1.0" encoding="UTF-8" standalone="yes"?><unit xmlns="http://www.srcML.org/srcML/src" revision="1.0.0" language="C" filename="test.c"><function><type><name>int</name></type> <name>main</name><parameter_list>()</parameter_list> <block>{<block_content><expr_stmt><expr><call><name>printf</name><argument_list>(<argument><expr><literal type="string">"Hello World!"</literal></expr></argument>)</argument_list></call></expr>;</expr_stmt><return>return <expr><literal type="number">0</literal></expr>;</return></block_content>}</block></function></unit>
'''
example_code = '''
int main() {
    printf("Hello World!");
    return 0;
}
'''

[INST] You have know how to generate the code. Now, here\'s the example code {example_code} and the srcml of that code {example}. can you follow the example to generate the srcml of the {code} without explanation -- just srcml [\INST]

\end{lstlisting}
}
\end{framed}

\textbf{Template for AST generation after finetuning with code translation} We used the following prompt for AST generation after finetuning with code translation, where \{code\}
is the function for AST generation. \{example\_c-
ode\} is the source code example for generating AST and the \{example\} is the AST for the \{example\_code\}.
\begin{framed}
\noindent
{\fontsize{9}{10}\selectfont
\begin{lstlisting}[
basicstyle=\ttfamily\fontsize{8}{9}\selectfont,
breaklines=true,
breakatwhitespace=false,
breakindent=0pt,
breakautoindent=false,
columns=fullflexible,
keepspaces=true,
aboveskip=0pt,
belowskip=0pt,
xleftmargin=0pt,
framexleftmargin=0pt
]
example = '''
<?xml version="1.0" encoding="UTF-8" standalone="yes"?><unit xmlns="http://www.srcML.org/srcML/src" revision="1.0.0" language="C" filename="test.c"><function><type><name>int</name></type> <name>main</name><parameter_list>()</parameter_list> <block>{<block_content><expr_stmt><expr><call><name>printf</name><argument_list>(<argument><expr><literal type="string">"Hello World!"</literal></expr></argument>)</argument_list></call></expr>;</expr_stmt><return>return <expr><literal type="number">0</literal></expr>;</return></block_content>}</block></function></unit>
'''
example_code = '''
int main() {
    printf("Hello World!");
    return 0;
}'''

[INST] You have know how to translate the code. Now, here\'s the example code {example_code} and the srcml of that code {example}. can you follow the example to generate the srcml of the {code} without explanation -- just srcml [\INST]

\end{lstlisting}
}
\end{framed}

\textbf{Template for callgraph generation after finetuning with code summarization} We used the following prompt for callgraph generation after finetuning with code summarization, where \{code\}
is the function for callgraph generation and \{method\_name\} is the name of the function.
\begin{framed}
\noindent
{\fontsize{9}{10}\selectfont
\begin{lstlisting}[
basicstyle=\ttfamily\fontsize{8}{9}\selectfont,
breaklines=true,
breakatwhitespace=false,
breakindent=0pt,
breakautoindent=false,
columns=fullflexible,
keepspaces=true,
aboveskip=0pt,
belowskip=0pt,
xleftmargin=0pt,
framexleftmargin=0pt
]
[INST] The Java code is {code}. You have known how to do the code summarization. Now, I want you to generate the callgraph path of the method {method_name} in the format of a->b->c etc, where -> means call and a is the method. Please follow the format and rule and do not give me the summary again. Remember the format of -> between each method name [\INST]

\end{lstlisting}
}
\end{framed}

\textbf{Template for callgraph generation after finetuning with code generation} We used the following prompt for callgraph generation after finetuning with code generation, where \{code\}
is the function for callgraph generation and \{method\_name\} is the name of the function.
\begin{framed}
\noindent
{\fontsize{9}{10}\selectfont
\begin{lstlisting}[
basicstyle=\ttfamily\fontsize{8}{9}\selectfont,
breaklines=true,
breakatwhitespace=false,
breakindent=0pt,
breakautoindent=false,
columns=fullflexible,
keepspaces=true,
aboveskip=0pt,
belowskip=0pt,
xleftmargin=0pt,
framexleftmargin=0pt
]
[INST]The Java code is {code}. You have known how to do the code generation. Now, I want you to generate the callgraph path of the method {method_name} in the format of a->b->c etc, where -> means call and a is the method. Please follow the format and rule and do not give me the code again. Remember the format of -> between each method name [\INST]

\end{lstlisting}
}
\end{framed}

\textbf{Template for callgraph generation after finetuning with code translation} We used the following prompt for callgraph generation after finetuning with code translation, where \{code\}
is the function for callgraph generation. \{method\_name\} is the name of the function.
\begin{framed}
\noindent
{\fontsize{9}{10}\selectfont
\begin{lstlisting}[
basicstyle=\ttfamily\fontsize{8}{9}\selectfont,
breaklines=true,
breakatwhitespace=false,
breakindent=0pt,
breakautoindent=false,
columns=fullflexible,
keepspaces=true,
aboveskip=0pt,
belowskip=0pt,
xleftmargin=0pt,
framexleftmargin=0pt
]
[INST]The Java code is {code}. You have known how to do the code translation. Now, I want you to generate the callgraph path of the method {method_name} in the format of a->b->c etc, where -> means call and a is the method. Please follow the format and rule and do not give me the code again. Remember the format of -> between each method name [\INST]

\end{lstlisting}
}
\end{framed}

\textbf{Template for dataflow generation after finetuning with code summarization} We used the following prompt for dataflow generation after finetuning with code summarization, where \{example\_data\_flow\} is the dataflow of the example code, \{example\_code\} is the code for the example dataflow,  \{code\}
is the function for dataflow generation, \{main\_method\_name\} is the source of the dataflow, and \{sink\} is the sink of the dataflow.
\begin{framed}
\noindent
{\fontsize{9}{10}\selectfont
\begin{lstlisting}[
basicstyle=\ttfamily\fontsize{8}{9}\selectfont,
breaklines=true,
breakatwhitespace=false,
breakindent=0pt,
breakautoindent=false,
columns=fullflexible,
keepspaces=true,
aboveskip=0pt,
belowskip=0pt,
xleftmargin=0pt,
framexleftmargin=0pt
]
example_code = '''
int main(int argc, char *argv[]) {
    if (argc > 1 && strcmp(argv[1], "42") == 0) {
        fprintf(stderr, "It depends!\n");
        exit(42);
    }
    printf("What is the meaning of life?\n");
    exit(0);
}                    
example_data_flow = '''
result: int main(int argc, char *argv[]) { -> if (argc > 1 && strcmp(argv[1], "42") == 0) { '''

[INST] You have known how to do code summarization. Now, I do not want you to generate the summary. Instead, I want you to generate the data flow of the provided code. Here is one example that you can follow {example_code} and its dataflow {example_data_flow}. could you please generate the data flow of the C code {code} from source {main_method_name} to sink {sink} in the format of a->b->c. Please follow the format and the example and do not give me the description [\INST]

\end{lstlisting}
}
\end{framed}

\textbf{Template for dataflow generation after finetuning with code generation} We used the following prompt for dataflow generation after finetuning with code generation, where \{example\_data\_flow\} is the dataflow of the example code, \{example\_code\} is the code for the example dataflow,  \{code\}
is the function for dataflow generation, \{main\_method\_name\} is the source of the dataflow, and \{sink\} is the sink of the dataflow.
\begin{framed}
\noindent
{\fontsize{9}{10}\selectfont
\begin{lstlisting}[
basicstyle=\ttfamily\fontsize{8}{9}\selectfont,
breaklines=true,
breakatwhitespace=false,
breakindent=0pt,
breakautoindent=false,
columns=fullflexible,
keepspaces=true,
aboveskip=0pt,
belowskip=0pt,
xleftmargin=0pt,
framexleftmargin=0pt
]
example_code = '''
int main(int argc, char *argv[]) {
    if (argc > 1 && strcmp(argv[1], "42") == 0) {
        fprintf(stderr, "It depends!\n");
        exit(42);
    }
    printf("What is the meaning of life?\n");
    exit(0);
}                     
example_data_flow = '''
result: int main(int argc, char *argv[]) { -> if (argc > 1 && strcmp(argv[1], "42") == 0) { '''
[INST] You have known how to do code generation. Now, I do not want you to generate the code. Instead, I want you to generate the data flow of the provided code. Here is one example that you can follow {example_code} and its dataflow {example_data_flow}. could you please generate the data flow of the C code {code} from source {main_method_name} to sink {sink} in the format of a->b->c. Please follow the format and the example and do not give me the description [\INST]

\end{lstlisting}
}
\end{framed}

\textbf{Template for dataflow generation after finetuning with code translation} We used the following prompt for dataflow generation after finetuning with code translation, where \{example\_data\_flow\} is the dataflow of the example code, \{example\_code\} is the code for the example dataflow,  \{code\}
is the function for dataflow generation, \{main\_method\_name\} is the source of the dataflow, and \{sink\} is the sink of the dataflow.
\begin{framed}
\noindent
{\fontsize{9}{10}\selectfont
\begin{lstlisting}[
basicstyle=\ttfamily\fontsize{8}{9}\selectfont,
breaklines=true,
breakatwhitespace=false,
breakindent=0pt,
breakautoindent=false,
columns=fullflexible,
keepspaces=true,
aboveskip=0pt,
belowskip=0pt,
xleftmargin=0pt,
framexleftmargin=0pt
]
example_code = '''
int main(int argc, char *argv[]) {
    if (argc > 1 && strcmp(argv[1], "42") == 0) {
        fprintf(stderr, "It depends!\n");
        exit(42);
    }
    printf("What is the meaning of life?\n");
    exit(0);
}                      
example_data_flow = '''
result: int main(int argc, char *argv[]) { -> if (argc > 1 && strcmp(argv[1], "42") == 0) {
'''

[INST] You have known how to do code translation. Now, I do not want you to translate the code. Instead, I want you to generate the data flow of the provided code. Here is one example that you can follow {example_code} and its dataflow {example_data_flow}. could you please generate the data flow of the C code {code} from source {main_method_name} to sink {sink} in the format of a->b->c. Please follow the format and the example and do not give me the description [\INST]

\end{lstlisting}
}
\end{framed}

\subsection{Research Methods -- RQ1}
\label{sec:rq2researchmethod}
Our research method for answering RQ1 is to evaluate closed-source LLMs on static analysis tasks without any finetuning. We compare the results with the tools that we describe in Section~\ref{sec:statictasks}. Our goal is to understand model capabilities rather than optimizing performance details. To this end, we used the default parameter settings and simple prompts throughout the entire dataset although different parameters might have slightly different results. As an additional experiment, we followed~\cite{kojima2022large} to add ``Let's think step by step`` at the end of our in-context learning prompt.~\cite{kojima2022large} observed that we can simply add ``Let's think step by step`` at the end of the prompt to improve the models' performance on reasoning tasks.

For C/C++, we used the testset that we extracted to for evaluation. For Java, we used the testset proposed by~\citet{leclair2019recommendations} for evaluation. For srcML, we used Levenshtein based on~\citep{decker2020srcdiff}. We used Jaccard similarity as our evaluation metrics for callgraph and data flow graph analysis~\citep{gharibi2018code2graph, blokhin2013malware, kargén2016towards}. We also computed the pair accuracy and chain accuracy for both tasks.

\subsection{Research Methods -- RQ2}
\label{sec:rq3researchmethod}
Our research method for answering RQ2 is to finetune the open-source LLMs and evaluate those models on static analysis tasks. We show those configurations in Tables~\ref{tab:datasets_models}~and~\ref{tab:rqconfigs}. For example, we show that we finetune \texttt{jam} to do ast generation and evaluate \texttt{jam} with the same task. We do not finetune \texttt{jam} for callgraph generation because of the context length limit (see Section~\ref{sec:codemodels}). Note that we increase the number of tokens to 10k for callgraph generation because callgraph generation would require larger context size.  We use the hyperparameters recommended by the papers accompanying each model, which we show in Table~\ref{tab:hyperparams}.  Recall from Section~\ref{sec:codemodels} that we used a QLoRA process for CodeLlaMA due to its size, but a full finetune process for jam. 

\subsection{Research Methods -- RQ3} 
Our research method for answering RQ3 is to finetune the open-source LLMs on downstream tasks after those models have been finetuned on static analysis tasks.  We show those configurations in Tables~\ref{tab:datasets_models}~and~\ref{tab:rqconfigs}. For example, in the first row of jam column in RQ3, we show that we first finetune \texttt{jam} on the AST generation task. Then, we further finetuned it for source code summarization. We compared this model with the one without finetuning with static analysis tasks to show whether static analysis improves the performance of LLMs on code intelligence tasks. 

The first row of Table~\ref{tab:datasets_models} show that the dataset we use for code summarization in Java is 170k dataset in~\cite{su2024distilled}.  However, Table~\ref{tab:datasets_models} also shows that we did not use jam for C/C++ datasets because \texttt{jam} is pretrained for Java only (see Section~\ref{sec:codemodels}).  Our guiding principle in all fine-tuning and evaluation is the same as RQ2.  We used the base jam and CodeLlaMa models released by their respective papers where those papers pretrained using large datasets of code (see Section~\ref{sec:codemodels}).  Then, we further trained those models with the training sets which were released with each dataset (see Sections~\ref{sec:devtasks}~and~\ref{sec:statictasks}).  We use the hyperparameters as described in RQ2.

We evaluated each configuration using the metrics and test datasets in the papers for each dataset.  For code summarization, we used the metrics METEOR and USE as implemented by~\citet{su2024distilled} and recommended by~\citet{haque2022semantic}.  For code translation and code generation, we used Bleu score~\citep{papineni2002bleu} as in the dataset that we used for experiments~\citep{iyer2018mapping,lu2021codexglue}. We also have additional datasets for code generation and translation evaluated with pass@k~\cite{chen2021evaluating} and compilation success rate.

\subsection{Research Methods -- RQ4}
Our research method for answering RQ4 is to finetune models with code intelligence tasks and evaluate the models with static anlysis tasks. To this end, we used CodeLlama fintuned with each code intelligence to generate the three different static analysis tasks. We show the configuration in Table~\ref{tab:rqconfigs}. For example, in the first row of RQ4, we show that we finetuned CodeLlama with code summarization. Then, we used the same model to generate AST. Note that we used simple prompts because our goal is to test the model's capability instead of finding the best results with better prompts.

\subsection{Research Methods -- RQ5}
Our research method for answering RQ5 is to classify the types of error  for static analysis tasks based on the current literature. To this end, we classify the error types for callgraph into four categories, which are missing direct call, extra direct calls, missing indirect calls, reference cycle, and predicted cycle based on~\cite{lehmann2023that,sui2020recall}.~\cite{lehmann2023that} constructed the callgraph that could potentially include cycles, direct edges, and indirect edges. We further classify into extra/missing direct calls and missing indirect calls based on false positive and false negative defined by~\cite{murphy1998empirical}, where false positive means additional edges and false negative means missing edges. For dataflow analysis, we followed~\cite{weideman2025data} to categorize the errors into direct edges and indirect edges. Based on false positive and false negative, we further classified these two edges into extra direct edges, extra indirect edges, missing direct edges, and missing indirect edges.  We categorized the AST errors into missing nodes and extra nodes based on AST difference discussed by~\cite{falleri2014finegrained, fan2021differential}. In addition,~\cite{alikhanifard2025novel} showed that AST can be incorrect semantically, so we added tag mismatch into our categories. We also added parsing error because a generated AST may have an incorrect format/structure that causes the compile errors.

\vspace{-4mm}

\subsection{Threats to Validity}
We organize threats to validity into threats to internal validity, external validity, and construct validity based on~\citet{wohlin2012experimentation}. The threats to internal validity are that different hyperparamters can have different results. We mitigate this threat by following the accompanying papers. The threats to construct validity are that our experimental results may not be able to generalize to different tasks and programming languages. We mitigate this by using three different code development tasks and static analysis tasks along with two different programming languages although more programming languages can have better generalizability of the findings. We leave the experiments for other programming languages as future work. The threats to external validity are that different models might generate different results. We mitigate this threat by using two different commercial LLMs and two different open-source LLMs.

\vspace{-4mm}

\subsection{Limitations}

Our prompt design is the key limitation in this paper. We used the simple prompt to test the reasoning capabilities of LLMs. The results show that LLMs do not do well on static analysis tasks. However, these results may not be able to generalize to more advanced prompting strategies (e.g., agentic development) and reasoning models, such as GPT-o series. Note that we did not have access to GPT-o series models at the time we conducted this experiment although~\cite{xie2025core} show that more advanced models still struggle with tasks that require deeper semantic understanding.

\section{Experimental Results}
In this section, we discuss our experimental results to each of our RQs.
\begin{table}[h!]
\centering
\small
\caption{Fine-tuning hyperparameters in our experiments.}
\label{tab:hyperparams}
\vspace{-2mm}
\begin{tabular}{lrr}
                                           & jam      & codellama \\ \cline{2-3} 
\multicolumn{1}{l|}{epoch}                 & 3        & 3         \\
\multicolumn{1}{l|}{learning rate}         & 3.00E-05 & 1.00E-04  \\
\multicolumn{1}{l|}{batch size}            & 4        & 2         \\
\multicolumn{1}{l|}{gradient accum. steps} & 32       & 16        \\
\multicolumn{1}{l|}{lora\_r}               & -        & 64        \\
\multicolumn{1}{l|}{lora\_alpha}           & -        & 16        \\
\multicolumn{1}{l|}{lora\_dropout}         & -        & 0         \\
\multicolumn{1}{l|}{quant\_type}           & -        & nf4       \\
\multicolumn{1}{l|}{bits}                  & -        & 4        
\end{tabular}
\end{table}



\begin{table}[h]
    \small
    \centering
    \caption{Results for srcml; gpt-base means the result without in-context learning; gpt-in-context means the result with in-context learning; codellama-finetuned means the result after finetuned.}
    \label{tab:srcml_result}
\begin{tabular}{llll} &   &     & \multicolumn{1}{c}{Metrics}                  \\
\multicolumn{1}{l|}{}       & task & \multicolumn{1}{c}{model} & \multicolumn{1}{c}{Levenshtein}                         \\ \hline

\multicolumn{1}{l|}{\multirow{7}{*}{Java}}                      & \multirow{7}{*}{srcml}  & \multicolumn{1}{c}{gpt-base} & \multicolumn{1}{c}{0.45}         \\
\multicolumn{1}{l|}{}                      &     & \multicolumn{1}{c}{gpt-in-context}                        & \multicolumn{1}{c}{0.80}      \\
\multicolumn{1}{l|}{}                      &    & \multicolumn{1}{c}{gemini-base}                            & \multicolumn{1}{c}{0.64}           \\ 
\multicolumn{1}{l|}{}                      &     & \multicolumn{1}{c}{gemini-in-context}                        & \multicolumn{1}{c}{0.93}       \\ 
\multicolumn{1}{l|}{}                      &     & \multicolumn{1}{c}{codellama-finetuned}                        & \multicolumn{1}{c}{0.77}         \\
\multicolumn{1}{l|}{}                      &     & \multicolumn{1}{c}{jam-finetuned}                        & \multicolumn{1}{c}{0.89}        \\
\hline
\multicolumn{1}{l|}{\multirow{7}{*}{C}}                      & \multirow{7}{*}{srcml}  & \multicolumn{1}{c}{gpt-base} & \multicolumn{1}{c}{0.46}         \\
\multicolumn{1}{l|}{}                      &     & \multicolumn{1}{c}{gpt-in-context}                        & \multicolumn{1}{c}{0.60}      \\
\multicolumn{1}{l|}{}                      &    & \multicolumn{1}{c}{gemini-base}                            & \multicolumn{1}{c}{0.65}           \\ 
\multicolumn{1}{l|}{}                      &     & \multicolumn{1}{c}{gemini-in-context}                        & \multicolumn{1}{c}{0.85}       \\ 
\multicolumn{1}{l|}{}                      &     & \multicolumn{1}{c}{codellama-finetuned}                        & \multicolumn{1}{c}{0.61}         \\
\multicolumn{1}{l|}{}                      &     & \multicolumn{1}{c}{jam-finetuned}                        & \multicolumn{1}{c}{-}        \\

\end{tabular}
\vspace{-3mm}
\end{table}

\begin{table*}[h]
\centering

\rotatebox{90}{%
\small
    \begin{tabular}{llllll}
    &   &     & \multicolumn{3}{c}{Metrics} \\
\multicolumn{1}{l|}{}       & task & \multicolumn{1}{c}{model} & \multicolumn{1}{c}{Jaccard similarity}      &  \multicolumn{1}{c}{pair accuracy}     &  \multicolumn{1}{c}{chain accuracy} \\ \hline

\multicolumn{1}{l|}{\multirow{13}{*}{Java}} & \multirow{6}{*}{callgraph}  & \multicolumn{1}{c}{gpt-base} & \multicolumn{1}{c}{0.121} & \multicolumn{1}{c}{0.123}  & \multicolumn{1}{c}{1.10E-04} \\
\multicolumn{1}{l|}{}                      &     & \multicolumn{1}{c}{gpt-in-context} & \multicolumn{1}{c}{0.122} & \multicolumn{1}{c}{0.123} & \multicolumn{1}{c}{6.45E-06} \\
\multicolumn{1}{l|}{}                      &     & \multicolumn{1}{c}{gpt-in-context-reason} & \multicolumn{1}{c}{0.113} & \multicolumn{1}{c}{0.114}    & \multicolumn{1}{c}{1.54E-05} \\
\multicolumn{1}{l|}{}                      &    & \multicolumn{1}{c}{gemini-base} & \multicolumn{1}{c}{0.186} & \multicolumn{1}{c}{0.190}   & \multicolumn{1}{c}{1.90E-04} \\ 
\multicolumn{1}{l|}{}                      &     & \multicolumn{1}{c}{gemini-in-context} & \multicolumn{1}{c}{0.191} & \multicolumn{1}{c}{0.195} & \multicolumn{1}{c}{1.20E-04} \\ 
\multicolumn{1}{l|}{}                      &     & \multicolumn{1}{c}{gemini-in-context-reason} & \multicolumn{1}{c}{0.176} & \multicolumn{1}{c}{0.180}    & \multicolumn{1}{c}{1.47E-07} \\

\multicolumn{1}{l|}{}                      &     & \multicolumn{1}{c}{codellama-finetuned} & \multicolumn{1}{c}{0.910} & \multicolumn{1}{c}{0.940}    & \multicolumn{1}{c}{2.20E-01} \\\cline{2-6}

\multicolumn{1}{l|}{}                      & \multirow{7}{*}{dataflow graph}  & \multicolumn{1}{c}{gpt-base} & \multicolumn{1}{c}{0.0047} & \multicolumn{1}{c}{ 0.0111}  & \multicolumn{1}{c}{0.0} \\
\multicolumn{1}{l|}{}                      &     & \multicolumn{1}{c}{gpt-in-context} & \multicolumn{1}{c}{0.0494} & \multicolumn{1}{c}{0.0951} & \multicolumn{1}{c}{5.37E-03} \\
\multicolumn{1}{l|}{}                      &     & \multicolumn{1}{c}{gpt-in-context-reason} & \multicolumn{1}{c}{0.0479} & \multicolumn{1}{c}{0.0941}    & \multicolumn{1}{c}{4.53E-03} \\
\multicolumn{1}{l|}{}                      &    & \multicolumn{1}{c}{gemini-base} & \multicolumn{1}{c}{0.0045} & \multicolumn{1}{c}{0.0114}   & \multicolumn{1}{c}{0.0} \\ 
\multicolumn{1}{l|}{}                      &     & \multicolumn{1}{c}{gemini-in-context} & \multicolumn{1}{c}{0.0522} & \multicolumn{1}{c}{0.1028}  & \multicolumn{1}{c}{2.31E-02} \\ 
\multicolumn{1}{l|}{}                      &     & \multicolumn{1}{c}{gemini-in-context-reason} & \multicolumn{1}{c}{0.0711} & \multicolumn{1}{c}{0.1426}    & \multicolumn{1}{c}{3.58E-03} \\
\multicolumn{1}{l|}{}                      &     & \multicolumn{1}{c}{codellama-finetuned} & \multicolumn{1}{c}{0.54} & \multicolumn{1}{c}{0.74} & \multicolumn{1}{c}{2.03E-01} \\
\multicolumn{1}{l|}{}                      &     & \multicolumn{1}{c}{jam-finetuned} & \multicolumn{1}{c}{0.38} & \multicolumn{1}{c}{0.54} & \multicolumn{1}{c}{0.18} \\\hline

\multicolumn{1}{l|}{\multirow{13}{*}{C}} & \multirow{6}{*}{callgraph}  & \multicolumn{1}{c}{gpt-base} & \multicolumn{1}{c}{0.160} & \multicolumn{1}{c}{0.200}  & \multicolumn{1}{c}{8.00E-06} \\
\multicolumn{1}{l|}{}                      &     & \multicolumn{1}{c}{gpt-in-context} & \multicolumn{1}{c}{0.217} & \multicolumn{1}{c}{0.280} & \multicolumn{1}{c}{0} \\
\multicolumn{1}{l|}{}                      &     & \multicolumn{1}{c}{gpt-in-context-reason} & \multicolumn{1}{c}{0.201} & \multicolumn{1}{c}{0.259}    & \multicolumn{1}{c}{1.69E-30} \\
\multicolumn{1}{l|}{}                      &    & \multicolumn{1}{c}{gemini-base} & \multicolumn{1}{c}{0.180} & \multicolumn{1}{c}{0.210}   & \multicolumn{1}{c}{6.47E-05} \\ 
\multicolumn{1}{l|}{}                      &     & \multicolumn{1}{c}{gemini-in-context} & \multicolumn{1}{c}{0.265} & \multicolumn{1}{c}{0.312} & \multicolumn{1}{c}{3.56E-05} \\ 
\multicolumn{1}{l|}{}                      &     & \multicolumn{1}{c}{gemini-in-context-reason} & \multicolumn{1}{c}{0.263} & \multicolumn{1}{c}{0.368}    & \multicolumn{1}{c}{1.20E-04} \\
\multicolumn{1}{l|}{}                      &     & \multicolumn{1}{c}{codellama-finetuned} & \multicolumn{1}{c}{0.310} & \multicolumn{1}{c}{0.550}    & \multicolumn{1}{c}{1.23E-04} \\\cline{2-6}

\multicolumn{1}{l|}{}                      & \multirow{7}{*}{dataflow graph}  & \multicolumn{1}{c}{gpt-base} & \multicolumn{1}{c}{0.0017} & \multicolumn{1}{c}{0.0054}  & \multicolumn{1}{c}{1.12E-04} \\
\multicolumn{1}{l|}{}                      &     & \multicolumn{1}{c}{gpt-in-context} & \multicolumn{1}{c}{0.0059} & \multicolumn{1}{c}{0.0126} & \multicolumn{1}{c}{5.13E-03} \\
\multicolumn{1}{l|}{}                      &     & \multicolumn{1}{c}{gpt-in-context-reason} & \multicolumn{1}{c}{0.0047} & \multicolumn{1}{c}{0.0118}    & \multicolumn{1}{c}{3.91E-03} \\

\multicolumn{1}{l|}{}                      &    & \multicolumn{1}{c}{gemini-base} & \multicolumn{1}{c}{0.0022} & \multicolumn{1}{c}{0.0056}   & \multicolumn{1}{c}{1.59E-03} \\ 
\multicolumn{1}{l|}{}                      &     & \multicolumn{1}{c}{gemini-in-context} & \multicolumn{1}{c}{0.0093} & \multicolumn{1}{c}{0.0176}  & \multicolumn{1}{c}{9.62E-03} \\ 
\multicolumn{1}{l|}{}                      &     & \multicolumn{1}{c}{gemini-in-context-reason} & \multicolumn{1}{c}{0.0047} & \multicolumn{1}{c}{0.0118}    & \multicolumn{1}{c}{3.91E-03} \\
\multicolumn{1}{l|}{}                      &     & \multicolumn{1}{c}{codellama-finetuned} & \multicolumn{1}{c}{0.12} & \multicolumn{1}{c}{0.25}     & \multicolumn{1}{c}{1.02E-01} \\
\multicolumn{1}{l|}{}                      &     & \multicolumn{1}{c}{jam-finetuned} & \multicolumn{1}{c}{-} & \multicolumn{1}{c}{-} & \multicolumn{1}{c}{-} 
    \end{tabular}
}
\caption{Results for callgraph and dataflow graph analysis.}
\label{tab:callgraph_datagraph_result}
\end{table*}

\begin{figure}[htbp]
    \centering
    \begin{subfigure}[b]{0.65\linewidth}
        \centering
        \includegraphics[width=\linewidth]{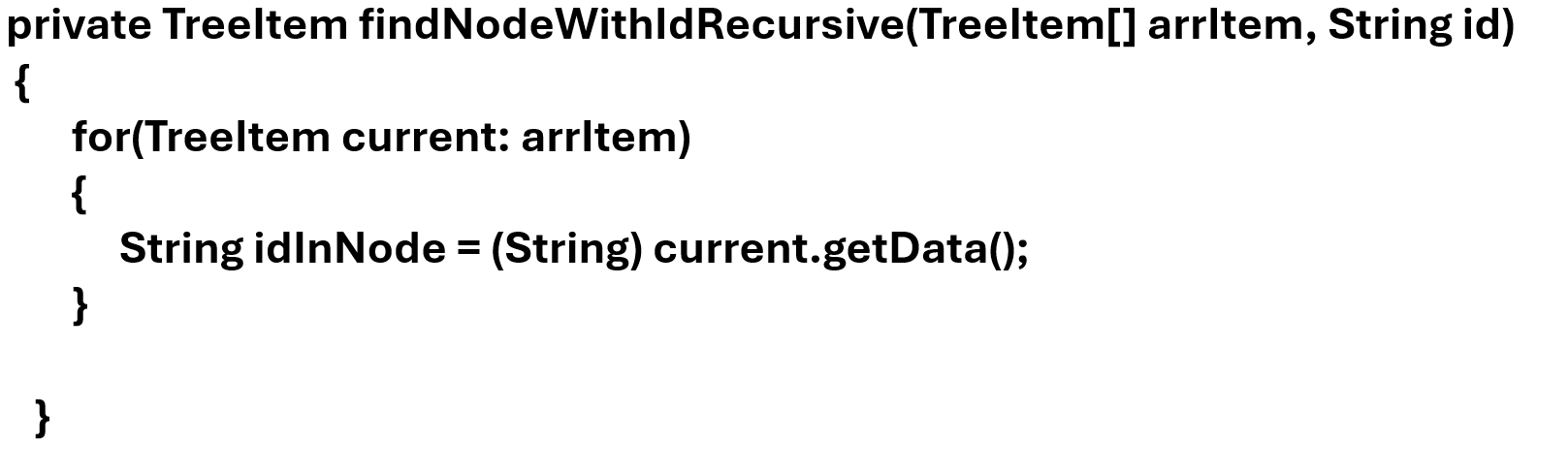}
        \caption{Source code}
        \label{fig:example_code}
    \end{subfigure}
    \vspace{2em}
    \begin{subfigure}[b]{0.65\linewidth}
        \centering
        \includegraphics[width=\linewidth]{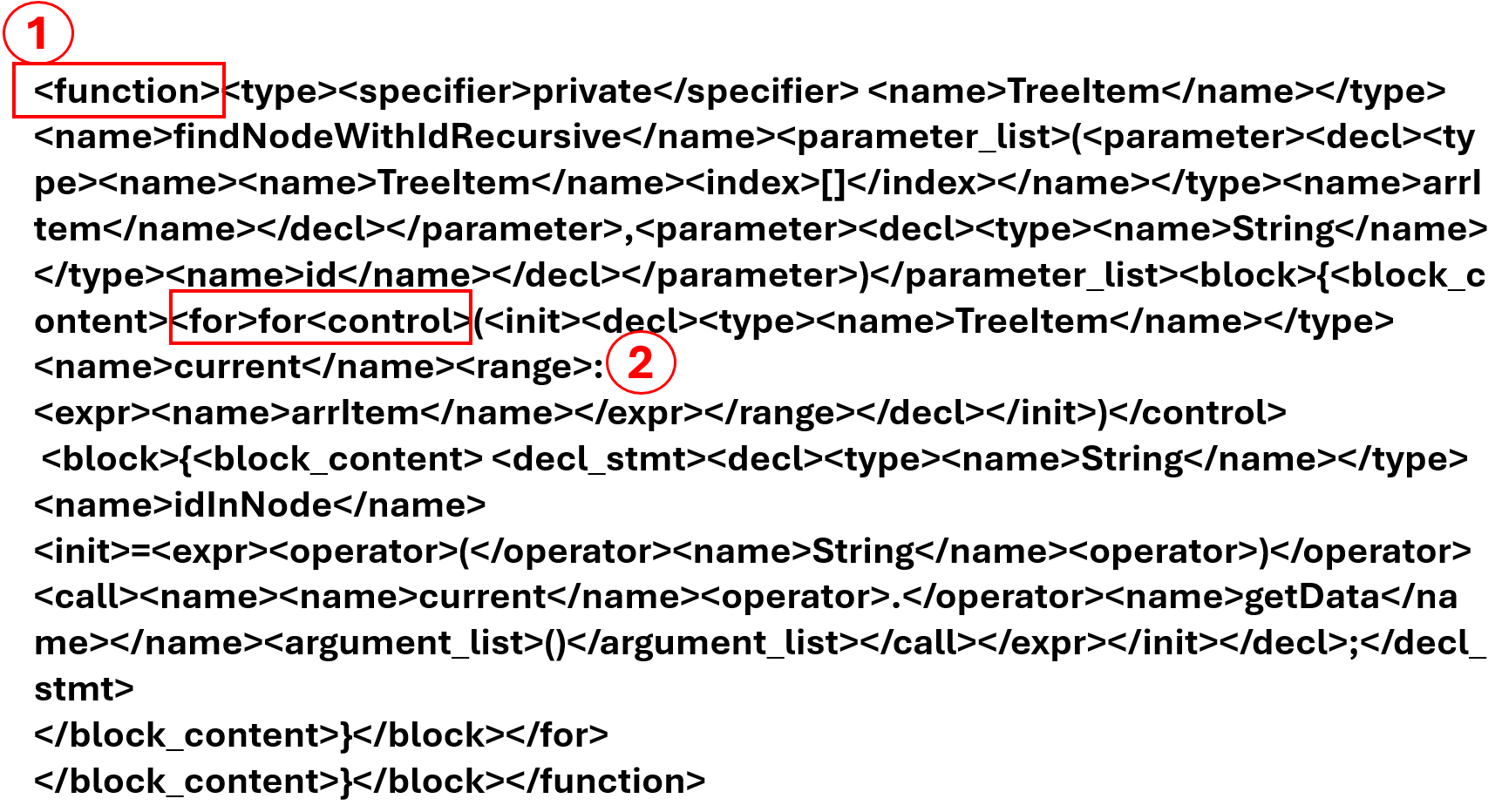}
        \caption{SrcML generated by tool}
        \label{fig:srcml_tool}
    \end{subfigure}
    \vspace{2em}
    \begin{subfigure}[b]{0.65\linewidth}
        \centering
        \includegraphics[width=\linewidth]{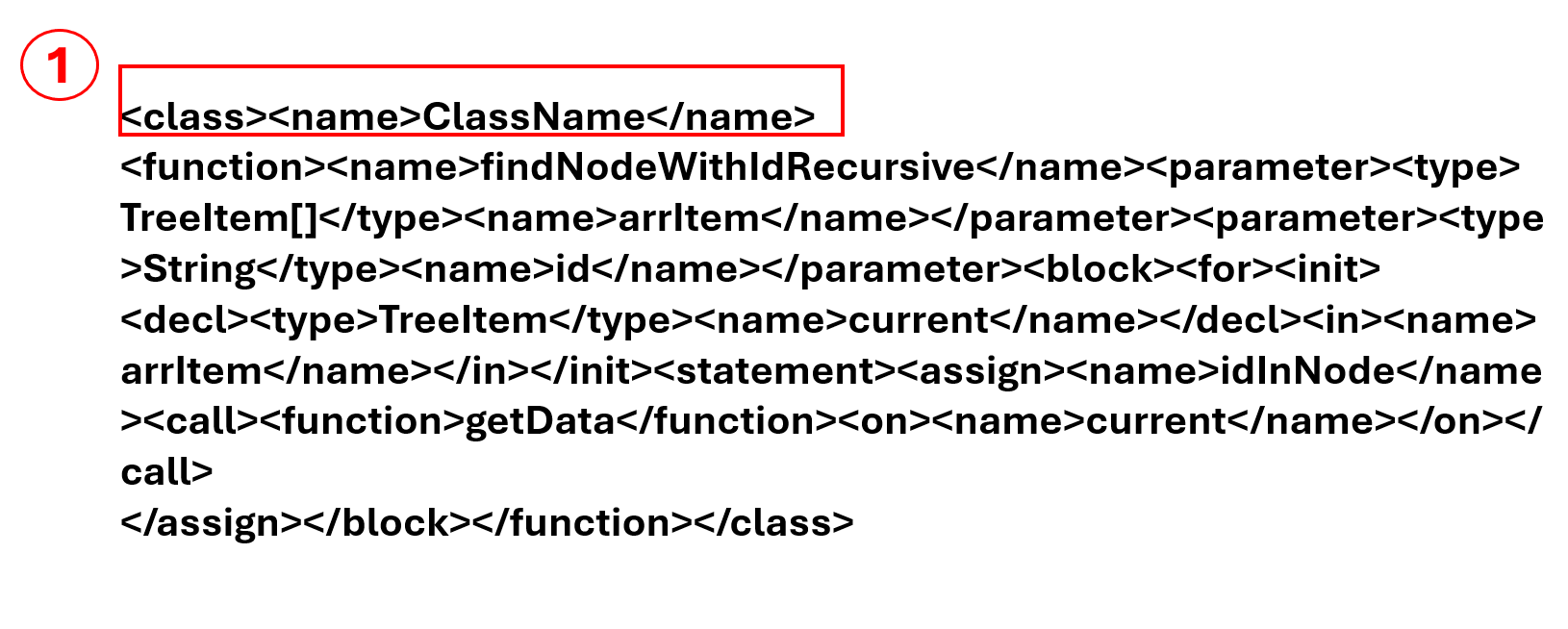}
        \caption{SrcML generated by GPT without in-context learning}
        \label{fig:srcml_gpt_witout_in_context}
    \end{subfigure}
    \vspace{2em}
    \begin{subfigure}[b]{0.65\linewidth}
        \centering
        \includegraphics[width=\linewidth]{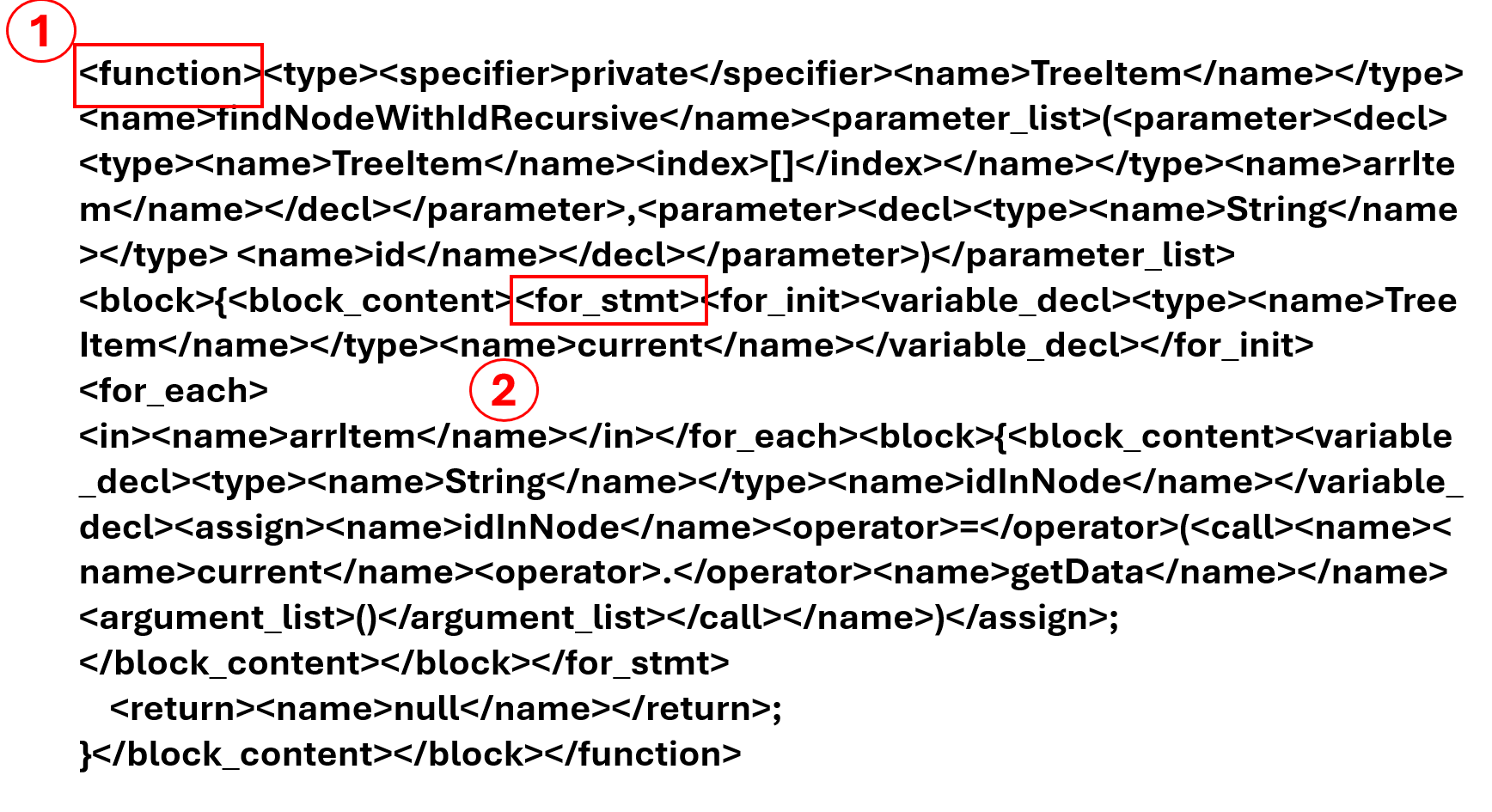}
        \caption{SrcML generated by GPT with in-context learning}
        \label{fig:srcml_gpt_context}
    \end{subfigure}
    
    \caption{Example of SrcML generated by GPT and tool for Java}
    \label{fig:srcmlgptex}
\end{figure}

\subsection{RQ1: Closed Models, Static Analysis Tasks}

We found that in-context learning improves AST generation tasks and popular programming languages have higher improvement. We showed the results for AST generation on both Java and C/C++ in Table~\ref{tab:srcml_result}. In Java, we found that \texttt{gpt-base} has the Levenshtein score of 0.45 and \texttt{gemini-base} has the Levenshtein score of 0.8. We also found the similar results in \texttt{gemini} although we found higher results in \texttt{gemini}, which has Levenshtein score of 0.64 for \texttt{gemini-base} and Levenshtein score of 0.93 for \texttt{gemini-in-context}. For C/C++, we observed the similar trend as in Java. However, we observed the smaller improvement in C/C++ dataset. For example, we found around 80\% improvement between GPT-base and GPT-in-context in Java dataset. However, we only found around 30\% improvement between GPT-base and GPT-in-context in C/C++ dataset, which has around 50\% difference in improvement. This result aligns with the finding that LLMs can have better performance on more popular programming languages~\citep{jimenez2024evaluation}. The gap between C/C++ and Java may imply the lack of reasoning skills on LLMs to generalize from one programming language to another programming language.

We showed an example from GPT in Figure~\ref{fig:srcmlgptex} for Java. We found that without in-context learning, the model tends to start the AST with \texttt{<class><name>} \texttt{ClassName</name>} even if we only use a single method as we showed in Area 1 in Figure~\ref{fig:srcml_gpt_witout_in_context}. With in-context learning, the method can correctly generate the AST that starts with \texttt{<function>} as shown in Area 1 in Figure~\ref{fig:srcml_gpt_context}. A possible explanation is that LLMs have been trained on huge dataset that has complete class, so those models assume all Java methods are inside a Java class. It is also possible that LLMs just try to learn the correct tags instead of correct syntax as in Area 2 in Figure~\ref{fig:srcml_gpt_context} and~\ref{fig:srcml_tool}. 

We found that LLMs do a poor job on callgraph and dataflow graph generation and in-context learning does not improve the results. We showed the results for callgraph and dataflow graph in Table~\ref{tab:callgraph_datagraph_result} for both Java and C/C++. For callgraph generation in Java, we found that \texttt{gpt-base} has Jaccard similarity of 0.121 and \texttt{gpt-in-context} has Jaccard similarity of 0.122. \texttt{gemini-base} and \texttt{gemini-in-context} have slightly higher performance, which is 0.186 and 0.191 respectively. We also found that chain accuracy is the lowest among all other metrics. In C/C++, we found the similar trend that in-context learning does not help to improve the results and chain accuracy is the lowest among all other metrics. For data flow generation, we found the similar results on both Java and C/C++ dataset. This result indicates that LLMs do not do well on callgraph and dataflow graph generation. 

We found that reasoning prompts do not improve the performance for both callgraph and dataflow graph generation. We show the results for reasoning prompts in Table~\ref{tab:callgraph_datagraph_result}. For example, we observe in Java that~\texttt{gpt-in-context} has Jaccard similarity of 0.122, but we only observe 0.113 in~\texttt{gpt-in-context-rea-}
\texttt{son}. The possible explanations are 1) LLMs do not have the fundamental ability for callgraph and dataflow generation, 2) in-context learning is only helping models to learn the format instead of the knowledge to do a task~\citep{long2024does}, and 3) models still struggle to comprehend longer prompt although LLMs can take very long context~\citep{zamfirescu2023why,liu2024lost}. A possible improvement would be guiding LLMs with more detailed instructions on each task instead of providing examples.

Overall, we found that closed-source LLMs do not do static analysis very well. To our surprise, we found that in-context learning only helps to improve AST generation but not callgraph generation and dataflow generation. The possible explanation is that it is easy for LLMs to learn the tags for AST generation. However, LLMs struggle with looking at a larger context for callgraph generation and more complicated reasoning tasks. These results suggest that LLMs do not have the same reasoning skills as human programmers.

\begin{figure}[htbp]
    \centering
    \begin{subfigure}[b]{0.65\linewidth}
        \centering
        \includegraphics[width=\linewidth]{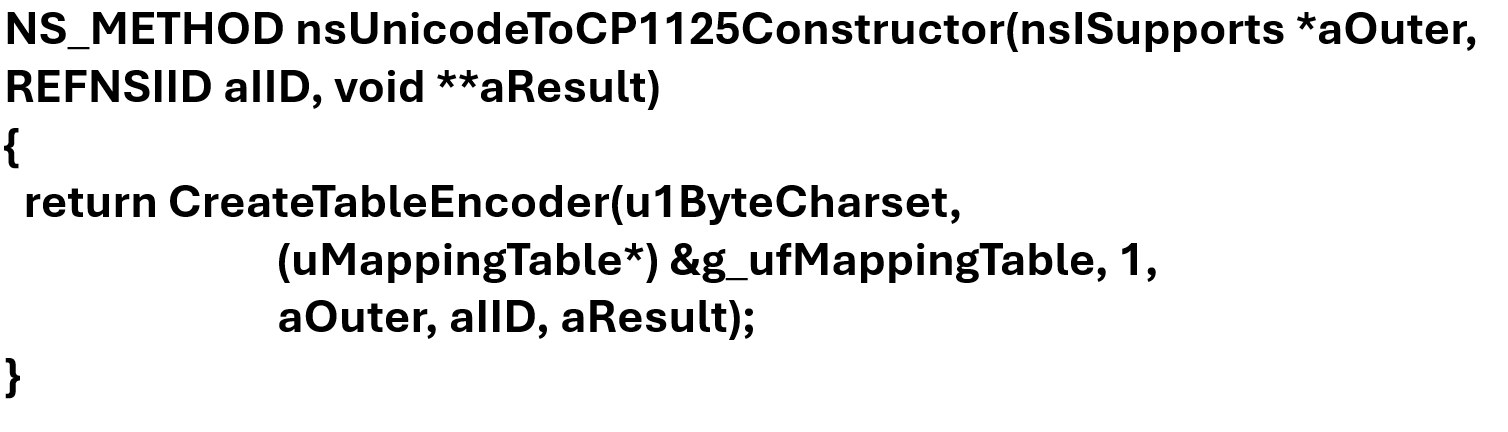}
        \caption{Source code}
        \label{fig:c_example_code}
    \end{subfigure}
    \vspace{2em}
    \begin{subfigure}[b]{0.65\linewidth}
        \centering
        \includegraphics[width=\linewidth]{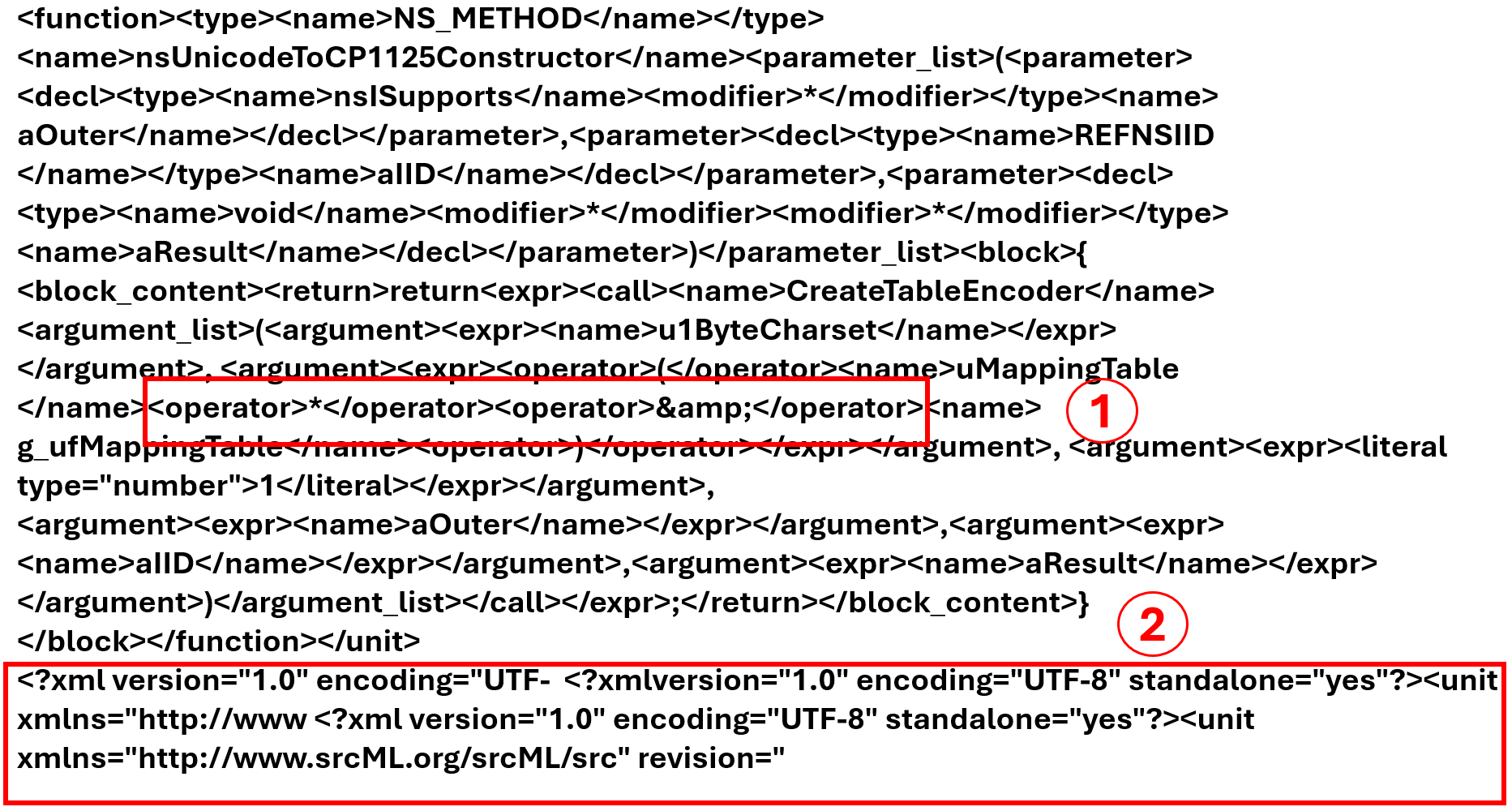}
        \caption{SrcML generated by CodeLlaMA}
        \label{fig:srcml_codellama_c}
    \end{subfigure}
    \vspace{2em}
    \begin{subfigure}[b]{0.65\linewidth}
        \centering
        \includegraphics[width=\linewidth]{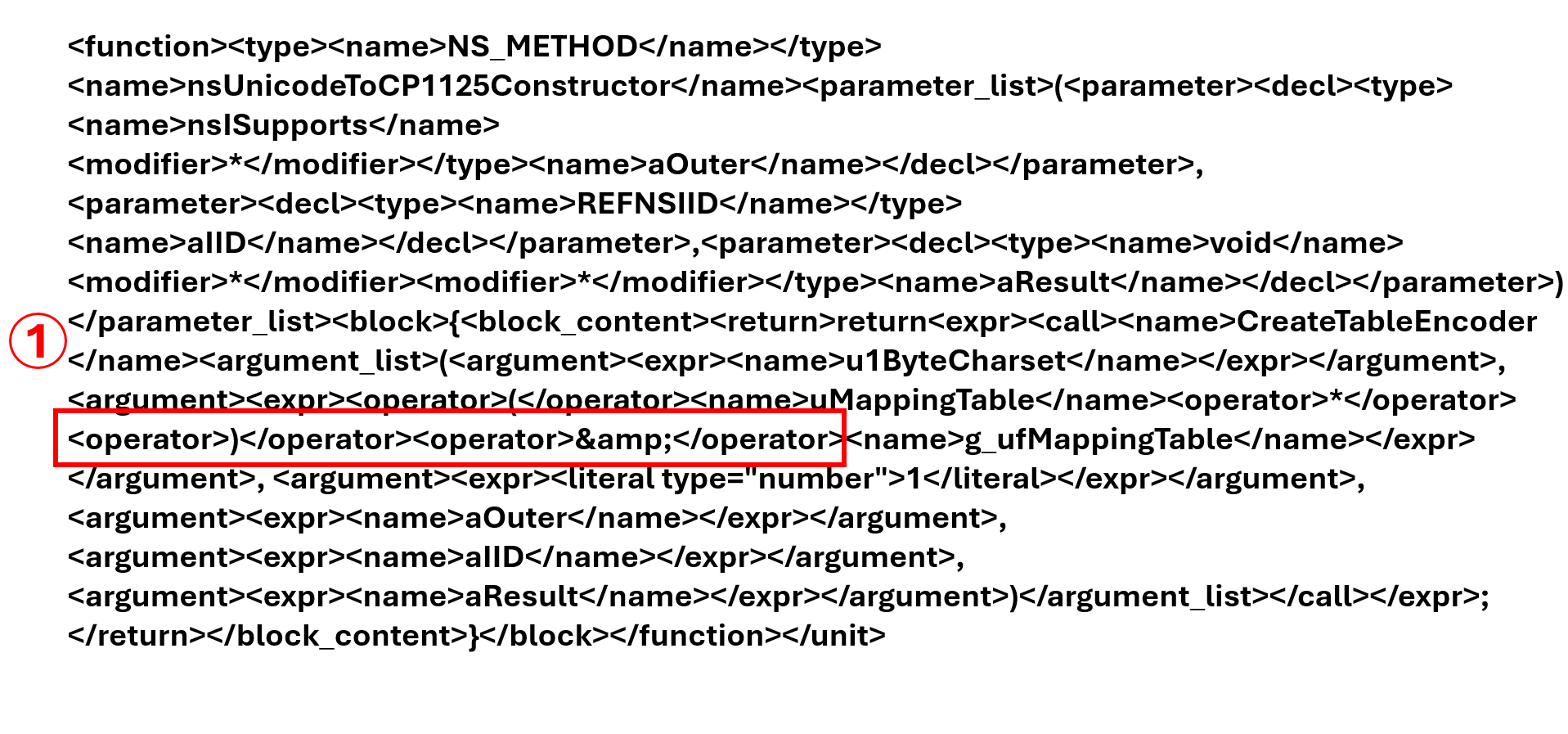}
        \caption{SrcML generated by tool}
        \label{fig:srcml_tool_c}
    \end{subfigure}
    
    \caption{Example of SrcML generated by CodeLlaMA and tool}
    \label{fig:srcmlcodellamaex}
\end{figure}

\vspace{-3mm}

\subsection{RQ2: Open Models, Static Analysis Tasks}
We observed that LLMs learn meaningful information after finetuning. Table~\ref{tab:srcml_result} shows the results for AST generation in Java and C/C++. In Java dataset, we observed that open-source models \texttt{jam} reached Levenshtein score of 0.89 and CodeLlaMA reached Levenshtein score of 0.77 for AST generation. We also found that the result of CodeLlaMA is slightly lower. The possible explanation is that we used less data to finetune the model due to the computational cost although CodeLlaMA is already a larger model with more pretrained data compared with \texttt{jam}. To our surprise, we found that \texttt{jam} even outperforms \texttt{gpt-in-context} in this task in the simple prompt settings. This shows that an appropriate finetuning helps to improve this task. For C/C++, we also observed the lower Levenshtein score compared with Java. Interestingly, we observed that LLMs can make a very simple syntactic error. For example, we observed in Figure~\ref{fig:srcml_codellama_c} in Area 1 that the AST generated by CodeLlaMA misses one ``)'' compared with Area 1 in Figure~\ref{fig:srcml_tool_c}. Moreover, we observed in Figure~\ref{fig:srcml_codellama_c} in Area 2 that LLMs might continue generating the token at the end. This shows the lack of semantic understanding in LLMs.

We observed that models can only partially generate the correct callgraph. We showed the result for both C/C++ and Java for callgraph generation in Table~\ref{tab:callgraph_datagraph_result}. We found that CodeLlaMA has high Jaccard similarity score and pair accuracy for callgraph generation after finetuning. However, we found that the chain accuracy is still very low. We also found that C/C++ dataset has lower score compared with Java dataset. The possible explanation is that we input the entire file for C/C++ dataset and this makes the problem harder. In addition, we only use 10k tokens during prediction due to hardware limitation. It is also possible that using entire files would require more tokens. However, the conclusion is that lower chain accuracy and higher pair accuracy and Jaccrad similarity would remain the same based on our findings from Java. This indicates that LLMs struggle to look at the longer context although the input size is long~\citep{li2024long}.

We observed that the results of dataflow generation are similar to callgraph generation, i.e., low chain accuracy but higher Jaccard similarity and pair accuracy and C/C++ dataset has lower results compared with Java. The results for both C/C++ and Java for dataflow graph generation are in Table~\ref{tab:callgraph_datagraph_result}. In Java, we also found that \texttt{jam} has very close results with CodeLlaMA on pair accuracy with the same finetuning data on data flow graph generation in Java dataset. This is to our surprise because larger models usually have bigger improvement in code development tasks, e.g., 25\% improvement on meteor score for code summarization in Java. However, it only shows a slight improvement or even worse on static analysis tasks. This result implies that LLMs saturate faster in static analysis tasks compared with other code intelligence tasks. A possible explanation is that it is harder for LLMs to learn the meaningful information in static analysis tasks.

To sum up, we found the similar results on both open-source models and closed-source models. Specifically, we found that LLMs have larger improvement on AST generation after finetuning. However, LLMs struggle with callgraph generation and dataflow graph generation. This result shows LLMs' inability on static analysis tasks.

\begin{table}[h]
\centering
\small
\vspace{-2mm}
\caption{Code summarization example for Java }
\label{tab:javacodesumex}
\begin{tabular}{ll}
\textbf Method ID 29591858 \\\hline
\end{tabular}

\begin{tabular}{c}
\begin{lstlisting}[language=Java, basicstyle=\ttfamily\small, breaklines=true, showstringspaces=false]
static public String makeName( List axes) {
    StringBuffer buff = new StringBuffer();
    for (int i=0; i<axes.size(); i++) {
      CoordinateAxis axis = (CoordinateAxis) axes.get(i);
      if (i>0) buff.append("-");
      buff.append( axis.getName());
    }
    return buff.toString();
  }




\end{lstlisting} \\

\end{tabular}

\begin{tabular}{l|p{6cm}}
\hline
     codellama-base & takes a list of CoordinateAxis objects and returns a string that concatenates their names separated by hyphens\\\hline
     codellama-srcml &  takes a list of CoordinateAxis objects and concatenates their names with hyphens in between, returning the resulting string \\\hline
     codellama-callgraph & takes a list of CoordinateAxis objects and concatenates their names with hyphens in between, returning the resulting string\\\hline
     codellama-dataflow & takes a list of CoordinateAxis objects and concatenates their names with hyphens in between, returning the resulting string \\\hline
     reference& takes a list of axes and returns a concatenated string of their names joined by "-"
\end{tabular}
\end{table}

\begin{table}[h]
\centering
\small
\vspace{-4.5mm}
\caption{Code summarization example for C/C++}
\label{tab:ccodesumex}
\begin{tabular}{ll}
\textbf Method ID 120 \\\hline
\end{tabular}

\begin{tabular}{c}
\begin{lstlisting}[language=C, basicstyle=\ttfamily\small, breaklines=true, showstringspaces=false]
static void knl_show_mc_route(u32 reg, char *s)
{
    int i;
    for (i = 0; i < KNL_MAX_CHANNELS; i++) {
        s[i*2] = knl_get_mc_route(i, reg) + '0';
        s[i*2+1] = '-';
    }
    s[KNL_MAX_CHANNELS*2 - 1] = '\0';
}
\end{lstlisting} \\ 
\end{tabular}

\begin{tabular}{l|p{6cm}}
\hline
     codellama-base & show the routing for a given MC channel \\\hline
     codellama-srcml &  show the multicast route for a given register \\\hline
     codellama-callgraph & show the routing for a given MC channel\\\hline
     codellama-dataflow & show the routing for a given MC channel \\\hline
     reference& Render the MC\_ROUTE register in human readable form
\end{tabular}
\end{table}

\begin{table}[!h]
\centering
\small
\caption{Code generation example for Java.}
\label{tab:javacodegenex}
\begin{tabular}{ll}
\textbf Method ID 988 \\\hline
\end{tabular}

\begin{tabular}{c}
\begin{lstlisting}[language=Java, basicstyle=\ttfamily\small, breaklines=true, showstringspaces=false]
boolean function(ProjCoordinate arg0, double arg1 ){ 
    double loc0 = arg0.x - this.x; 
    if ( loc0 > arg1 ) { return false; } 
    else { return true; } 
} 


\end{lstlisting} \\

\end{tabular}

\begin{tabular}{l|p{6cm}}
\hline
     jam-base &\begin{lstlisting}[language=Java, basicstyle=\ttfamily\scriptsize, breaklines=true, showstringspaces=false]
boolean function( double arg0){ 
return (( arg0 >= this.x ) && ( arg0 <= this.x + this.z) && ( ( arg0 >= this.y ) && ( arg0 <= this.


\end{lstlisting} \\\hline
     jam-srcml &  \begin{lstlisting}[language=Java, basicstyle=\ttfamily\scriptsize, breaklines=true, showstringspaces=false]
boolean function(double arg0, double arg1 ){ 
return (arg0 >= x) && (arg0 <= arg1); }


\end{lstlisting}\\\hline
     jam-dataflow &  \begin{lstlisting}[language=Java, basicstyle=\ttfamily\scriptsize, breaklines=true, showstringspaces=false]
boolean function(Coordinate arg0 ){ 
return ( x == arg0.x && y == arg0.y && z == arg0.z );
}


\end{lstlisting}  \\\hline
\end{tabular}
\end{table}

\begin{table}[h]
\centering
\small
\caption{Code generation example for C/C++.}
\label{tab:ccodegenex}
\begin{tabular}{ll}
\textbf Method ID 10 \\\hline
\end{tabular}

\begin{tabular}{c}
\begin{lstlisting}[language=C++, basicstyle=\ttfamily\small, breaklines=true, showstringspaces=false]

int binaryDec(int arr [], int n){ 
    int ans = 0; 
    for(int i = 0 ; i < n ; i ++ ) 
        ans += arr [ i ] << ( n - i - 1 ); 
    return ans; 
}

\end{lstlisting} \\

\end{tabular}

\begin{tabular}{l|p{7cm}}
\hline
codellama-base &\begin{lstlisting}[language=Java, basicstyle=\ttfamily\scriptsize, breaklines=true, showstringspaces=false]
int toDecimal(int arr[] , int n) { 
    int res = 0 ; 
    for(int i = 0; i < n;i ++){ 
        res += arr [i] * ( 1 << i ); 
    } 
    return res; 
} 

\end{lstlisting} \\\hline
     codellama-srcml &  \begin{lstlisting}[language=Java, basicstyle=\ttfamily\scriptsize, breaklines=true, showstringspaces=false]
int toDecimal(int arr [], int n){ 
    int res = 0; 
    for(int i = 0; i < n; i ++ ) 
        res += arr [ i ] * ( 1 << i ); 
    return res; 
} 


\end{lstlisting}\\\hline
     codellama-dataflow &  \begin{lstlisting}[language=Java, basicstyle=\ttfamily\scriptsize, breaklines=true, showstringspaces=false]
int toDecimal(int arr [], int n){ 
    int res = 0; 
    for(int i = 0; i < n; i ++){ 
        res += arr [i] * ( 1 << i ); 
    } 
    return res;
}
\end{lstlisting}\\\hline
codellama-callgraph &  \begin{lstlisting}[language=Java, basicstyle=\ttfamily\scriptsize, breaklines=true, showstringspaces=false]
int convertFromBinary(int arr []){ 
    int n = 0 , i = 0 ; 
    while(arr[i] != 0 ){ 
        n = n * 2 + arr [i]; 
        i ++; 
    } 
    return n; 
}
    

\end{lstlisting}  \\\hline
\end{tabular}
\end{table}

\begin{table}[h]
\centering
\small
\caption{Code generation example for C/C++.}
\label{tab:codegenerrorex}
\begin{tabular}{ll}
\textbf Method ID 10 \\\hline
\end{tabular}

\begin{tabular}{c}
\begin{lstlisting}[language=C++, basicstyle=\ttfamily\small, breaklines=true, showstringspaces=false]

int EnneacontahexagonNum(int n){ 
    return (94 * n * n - 92 * n) / 2; 
}

\end{lstlisting} \\

\end{tabular}

\begin{tabular}{l|p{7cm}}
\hline
codellama-base &\begin{lstlisting}[language=Java, basicstyle=\ttfamily\scriptsize, breaklines=true, showstringspaces=false]
int enneacontahexagonNum(int n){ 
     return ( 118 * n * n - 106 * n ) / 2; 
}

\end{lstlisting} \\\hline
     codellama-srcml &  \begin{lstlisting}[language=Java, basicstyle=\ttfamily\scriptsize, breaklines=true, showstringspaces=false]
int enneacontahexagonNum(int n){ 
    return ( 11 * n * n - 10 * n ) / 2; 
}


\end{lstlisting}\\\hline
     codellama-dataflow &  \begin{lstlisting}[language=Java, basicstyle=\ttfamily\scriptsize, breaklines=true, showstringspaces=false]
int enneacontahexagonNum(int n){ 
    return (11 * n * n - 10 * n) / 2; 
} 
\end{lstlisting}\\\hline
codellama-callgraph &  \begin{lstlisting}[language=Java, basicstyle=\ttfamily\scriptsize, breaklines=true, showstringspaces=false]
int enneacontahexagonNum(int n){ 
    return (14 * n * n - 14 * n + 1) / 2; 
}
    

\end{lstlisting}  \\\hline
\end{tabular}
\end{table}

\subsection{RQ3: Open Models, Static Analysis + Dev.}
We do not strongly find the statistical difference between the models pretrained with static analysis and the baselines on both programming languages (Java and C/C++) and both models (CodeLlama and Jam) in code summarization. We showed the results for code summarization in Table~\ref{tab:rq1_codesum_results} for Java and C/C++.  The possible explanation is that the model looks at the keyword in the method for source code summarization instead of using static analysis tasks as what human programmers would do to comprehend source code. We showed an example in Java in Table~\ref{tab:javacodesumex}. We found that LLMs tend to copy the technical jargon from the method for code summary. Specifically, we found that the summary from CodeLlaMA uses ``CoordinateAxis'' in the summary even after static analysis tasks. Similarly, we also observed the same phenomenon in C/C++. For example, in Table~\ref{tab:ccodesumex}, we found that human reference uses ``render'' as a verb to describe the method. However, the summary from CodeLlaMA uses ``show'' as a verb, which matches the signature of the function. Interestingly, we found that some of the summaries are the same. This is because we use very small temperature, i.e., 1e-9 and models may generate the same result when the weights are similar~\citep{peeperkorn2024temperature}, which further showed that pretraining models with static analysis tasks does not improve code summarization tasks significantly.

In terms of BLEU and CodeBERT, we do not observe consistent improvement when we pretrain the models with static analysis tasks before finetuning models with code generation. Table~\ref{tab:rq1_codegen_results} shows the result for code generation on both Java and C/C++. The possible explanation is that LLMs do not use static analysis tasks as the thought process. For example, in Table~\ref{tab:javacodegenex}, we found that code generated by \texttt{jam-srcml} has better syntactic structure compared with \texttt{jam-base}. However, the underlying logic is still different from the reference. Another example in C/C++ dataset in Table~\ref{tab:ccodegenex} is that we observed that the code generated by CodeLlaMA without pretraining with any static analysis tasks has the logical error in the ``toDecimal'' method. After pretraining with static analysis tasks, we still found the same logical error. Interestingly, we observed that LLMs might generate code with incorrect details. In Table~\ref{tab:codegenerrorex}, we found that CodeLlaMA does not generate the correct parameters for Enneacontahexagon numbers. For human programmers to do this task, they would look at the correct parameters before writing code. This indicates that static analysis tasks help LLMs better understand syntactic but not semantic information. Similar results are observed by~\cite{ma2023lms}. 

For execution-based evaluation, we observed similar results as evaluated with BLEU score and CodeBERT in code generation. We show the result of the execution-based evaluation for code generation in Table~\ref{tab:rq1_codegen_execution_results}.  We found that static analysis tasks improve the compilation success rate, but we do not observe consistent improvement over pass@1. The possible explanation is that static analysis tasks improve the syntax of source code, but the underlying logic is still wrong. Note that the dataset that we used is extremely difficult as discussed in the original paper~\citep{li2022competition}, which requires a large amount of samples to have a higher pass@k. Our goal in this paper is to understand the model capabilities instead of finding the optimal results. To this end, our goal is to ensure every model has the same settings instead of finding the optimal settings for each model.

For the code translation dataset, we observe statistical significant improvement in the BLEU score in the Java dataset after we pretrained models with callgraphs. However, we do not observe the same improvement in the C/C++ dataset. We show the results for both Java and C/C++ in Table~\ref{tab:rq1_codetrans_results}. The possible explanation is that pretraining models with callgraphs help models look at the larger context and code translation would require a larger context for more accurate translation. However, we cannot strongly attribute the improvement to pretrainig models with callgraphs because we do not observe similar improvement in the C to Rust dataset. Interestingly, we found that code translation has higher performance compared with other tasks in Java datasets. It is also possible that Java and C\# have very similar API calls and syntax, so LLMs can easily copy the information from the source language as we do not observe a similar trend in the C to Rust dataset.

Regarding execution-based metrics on code translation, we observed that callgraph improves the compilation success rate, but none of the static analysis tasks improves the pass@1. We show the results of the execution-based evaluation for code translation in Table~\ref{tab:rq1_codetrans_execution_results}. The possible explanation is that callgraph provides the more information between the function calls for code translation tasks. However, LLMs do not learn the semantic information during pretraining. This indicates the lack of semantic reasoning skills in LLMs. Note that our goal is to test the model's capabilities instead of finding the best performance, so we only used the simple prompts. The low pass@1 also aligns with the results in~\cite{pan2024lost} in llama-based models.

It is possible that we do not use proper prompts to guide the models to perform downstream tasks after pretraining. However,~\cite{lu2024emergent} observed that the emerging ability of LLMs comes from linguistic knowledge and model memory. To this end, we examine the performance of jam models more carefully because jam models require the specific prompt format instead of free-form, which reduces the threat of improper prompts. We show the results in Table~\ref{tab:rq1_codesum_results}, Table~\ref{tab:rq1_codegen_results}, and Table~\ref{tab:rq1_codetrans_results}. We do not observe strong improvement over three code intelligence tasks. Specifically, we do not observe any statistical difference for source code summarization and code generation. This shows that models do not use static analysis as the fundamental knowledge to perform code intelligence tasks as human programmers.

Overall, our results lead to two conclusions 1) LLMs might have an alien thought process for code comprehension compared with human programmers 2) the static analysis tasks improve the overall syntax of the code, but static analysis tasks do not improve the semantics.

\begin{table}[!h]
\centering
\small
\caption{Example that shows the common errors in Dataflow graph generation. The sink is PR\_Unlock and the source is from the function signature.}
\label{tab:dataflowqulatative}
\vspace{-1mm}
\begin{tabular}{ll}
\end{tabular}

\begin{tabular}{c}
\begin{lstlisting}[language=C, basicstyle=\ttfamily\small, breaklines=true, showstringspaces=false]

void jsd_Unlock(JSDStaticLock* lock){
    void* me;
    ASSERT_VALID_LOCK(lock);
    _CURRENT_THREAD(me);
    JS_ASSERT(lock->owner == me);
    if(lock->owner != me)
        return;
    if(--lock->count == 0)
    {
        lock->owner = NULL;
        PR_Unlock(lock->lock);
    }
}

\end{lstlisting} \\

\end{tabular}

\begin{tabular}{l|p{7cm}}
\hline
reference &\begin{lstlisting}[language=C, basicstyle=\ttfamily\scriptsize, breaklines=true, showstringspaces=false]
jsd_Unlock(JSDStaticLock* lock) => ASSERT_VALID_LOCK(lock); => PR_Unlock(lock->lock);

\end{lstlisting} \\\hline
     codellama &  \begin{lstlisting}[language=Java, basicstyle=\ttfamily\scriptsize, breaklines=true, showstringspaces=false]
1. _CURRENT_THREAD(me) => JS_ASSERT(lock->owner == me)
2. JS_ASSERT(lock->owner == me) => if(lock->owner != me)
3. if(lock->owner != me) => return
4. if(--lock->count == 0) => lock->owner = NULL
5. lock->owner = NULL => PR_Unlock(lock->lock)
6. if(--lock->count == 0) => PR_Unlock(lock->lock)


\end{lstlisting}\\\hline

\end{tabular}
\end{table}

\begin{table}[]
    \centering
    \small
    \caption{Meteor and USE score for code summarization}
    \label{tab:rq1_codesum_results}
\begin{tabular}{lllllll} &   &     & \multicolumn{2}{c}{Metrics}               &   \\
\multicolumn{1}{l|}{}       & model & \multicolumn{1}{c}{config} & \multicolumn{1}{c}{METEOR}      &  \multicolumn{1}{c}{USE}     &  \multicolumn{1}{c}{p\_m}    &  \multicolumn{1}{c}{p\_u}                         \\ \hline
\multicolumn{1}{l|}{\multirow{8}{*}{Java}}  & \multirow{4}{*}{jam}       & \multicolumn{1}{c}{base} &     \multicolumn{1}{c}{40.95}     &     \multicolumn{1}{c}{68.80}   &     \multicolumn{1}{c}{-}   &     \multicolumn{1}{c}{-}       \\
\multicolumn{1}{l|}{}                      &  &  \multicolumn{1}{c}{srcml}                        & \multicolumn{1}{c}{41.28} &     \multicolumn{1}{c}{69.01}     &     \multicolumn{1}{c}{0.02}     &     \multicolumn{1}{c}{0.06}         \\
\multicolumn{1}{l|}{}                      &    &  \multicolumn{1}{c}{dataflow graph}                             & \multicolumn{1}{c}{40.09} &    \multicolumn{1}{c}{68.32} &    \multicolumn{1}{c}{0.99} &       \multicolumn{1}{c}{0.99}      \\ 
\multicolumn{1}{l|}{}                      &                      &  \multicolumn{1}{c}{callgraph}      & \multicolumn{1}{c}{-} &   \multicolumn{1}{c}{-} & \multicolumn{1}{c}{-}     & \multicolumn{1}{c}{-}            \\ 
\cline{2-7} 

\multicolumn{1}{l|}{}                      & \multirow{4}{*}{codellama}  & \multicolumn{1}{c}{base} & \multicolumn{1}{c}{51.44} &      \multicolumn{1}{c}{74.56}  &      \multicolumn{1}{c}{-}  &      \multicolumn{1}{c}{-}         \\
\multicolumn{1}{l|}{}                      &     & \multicolumn{1}{c}{srcml}                        & \multicolumn{1}{c}{51.19} &    \multicolumn{1}{c}{74.48} &    \multicolumn{1}{c}{0.99}   &    \multicolumn{1}{c}{0.88}         \\
\multicolumn{1}{l|}{}                      &    & \multicolumn{1}{c}{dataflow graph}                            & \multicolumn{1}{c}{51.27} & \multicolumn{1}{c}{74.54}   &    \multicolumn{1}{c}{0.94}&    \multicolumn{1}{c}{0.62}         \\ 
\multicolumn{1}{l|}{}                      &     & \multicolumn{1}{c}{callgraph}                        & \multicolumn{1}{c}{51.43} &        \multicolumn{1}{c}{74.57} &    \multicolumn{1}{c}{0.53}&    \multicolumn{1}{c}{0.43}        \\\hline
\multicolumn{1}{l|}{\multirow{4}{*}{C}}  & \multirow{4}{*}{codellama}    & \multicolumn{1}{c}{base} & \multicolumn{1}{c}{29.65}   & \multicolumn{1}{c}{50.62} & \multicolumn{1}{c}{-}   & \multicolumn{1}{c}{-}   \\
\multicolumn{1}{l|}{}                      &                            & \multicolumn{1}{c}{srcml} & \multicolumn{1}{c}{29.96} & \multicolumn{1}{c}{50.85}& \multicolumn{1}{c}{0.99} & \multicolumn{1}{c}{0.99} \\
\multicolumn{1}{l|}{}                      &                            & \multicolumn{1}{c}{dataflow graph} & \multicolumn{1}{c}{29.87} & \multicolumn{1}{c}{50.73} & \multicolumn{1}{c}{0.99} & \multicolumn{1}{c}{0.93} \\
\multicolumn{1}{l|}{}                      &       & \multicolumn{1}{c}{callgraph}                      & \multicolumn{1}{c}{29.97} & \multicolumn{1}{c}{50.91}& \multicolumn{1}{c}{0.99}  & \multicolumn{1}{c}{0.99}  \\\hline

\end{tabular}
\vspace{-3mm}
\end{table}

\begin{table}[]
\vspace{3mm}
    \centering
    \small
    \caption{BLEU and CodeBERT  for code generation}
    \label{tab:rq1_codegen_results}
\begin{tabular}{lllllll} 
\multicolumn{1}{l|}{}       & model & \multicolumn{1}{c}{config} & \multicolumn{1}{c}{BLEU} & \multicolumn{1}{c}{CodeBERT} &  \multicolumn{1}{c}{p\_b} & \multicolumn{1}{c}{p\_c} \\ \hline
\multicolumn{1}{l|}{\multirow{8}{*}{Java}}  & \multirow{4}{*}{jam}       & \multicolumn{1}{c}{base} & \multicolumn{1}{c}{23.51} & \multicolumn{1}{c}{0.95} & \multicolumn{1}{c}{-} & \multicolumn{1}{c}{-} \\
\multicolumn{1}{l|}{}                      &  &  \multicolumn{1}{c}{srcml} & \multicolumn{1}{c}{24.00} & \multicolumn{1}{c}{0.95} & \multicolumn{1}{c}{$p<0.01$} & \multicolumn{1}{c}{$p<0.01$} \\
\multicolumn{1}{l|}{}                      &    &  \multicolumn{1}{c}{dataflow graph} & \multicolumn{1}{c}{22.79} & \multicolumn{1}{c}{0.95} & \multicolumn{1}{c}{0.63} & \multicolumn{1}{c}{0.30} \\ 
\multicolumn{1}{l|}{}                      &                      &  \multicolumn{1}{c}{callgraph} & \multicolumn{1}{c}{-} & \multicolumn{1}{c}{-} & \multicolumn{1}{c}{-} & \multicolumn{1}{c}{-} \\ 
\cline{2-7} 

\multicolumn{1}{l|}{}                      & \multirow{4}{*}{codellama}  & \multicolumn{1}{c}{base} & \multicolumn{1}{c}{31.42} & \multicolumn{1}{c}{0.97} & \multicolumn{1}{c}{-} & \multicolumn{1}{c}{-} \\
\multicolumn{1}{l|}{}                      &     & \multicolumn{1}{c}{srcml} & \multicolumn{1}{c}{30.40} & \multicolumn{1}{c}{0.97} & \multicolumn{1}{c}{0.92} & \multicolumn{1}{c}{$p<0.05$} \\
\multicolumn{1}{l|}{}                      &    & \multicolumn{1}{c}{dataflow graph} & \multicolumn{1}{c}{34.22} & \multicolumn{1}{c}{0.97} & \multicolumn{1}{c}{$p<0.001$} & \multicolumn{1}{c}{$p<0.001$} \\ 
\multicolumn{1}{l|}{}                      &     & \multicolumn{1}{c}{callgraph} & \multicolumn{1}{c}{28.93} & \multicolumn{1}{c}{0.97} & \multicolumn{1}{c}{0.99} & \multicolumn{1}{c}{0.62} \\\hline

\multicolumn{1}{l|}{\multirow{4}{*}{C}}  & \multirow{4}{*}{codellama}    & \multicolumn{1}{c}{base} & \multicolumn{1}{c}{48.20} & \multicolumn{1}{c}{0.93} & \multicolumn{1}{c}{-} & \multicolumn{1}{c}{-} \\
\multicolumn{1}{l|}{}                      &                          & \multicolumn{1}{c}{srcml} & \multicolumn{1}{c}{48.18} & \multicolumn{1}{c}{0.93} & \multicolumn{1}{c}{0.70} & \multicolumn{1}{c}{0.38} \\
\multicolumn{1}{l|}{}                      &                            & \multicolumn{1}{c}{dataflow graph} & \multicolumn{1}{c}{47.54} & \multicolumn{1}{c}{0.93} & \multicolumn{1}{c}{0.50} & \multicolumn{1}{c}{0.53} \\
\multicolumn{1}{l|}{}                      &       & \multicolumn{1}{c}{callgraph} & \multicolumn{1}{c}{47.82} & \multicolumn{1}{c}{0.93} & \multicolumn{1}{c}{0.73} & \multicolumn{1}{c}{0.76} \\\hline 

\end{tabular}
\vspace{-3mm}
\end{table}

\begin{table}[]
\vspace{3mm}
    \centering
    \small
    \caption{Execution-based evaluation on code generation}
    \label{tab:rq1_codegen_execution_results}
\begin{tabular}{lllllll}
\multicolumn{1}{l|}{}       
& model 
& \multicolumn{1}{c}{config} 
& \multicolumn{1}{c}{Pass@1} 
& \multicolumn{1}{c}{Compilation rate}  
\\ \hline

\multicolumn{1}{l|}{\multirow{4}{*}{C}}                  
& \multirow{4}{*}{codellama}  
& \multicolumn{1}{c}{base} 
& \multicolumn{1}{c}{0.024} 
& \multicolumn{1}{c}{0.45} 
\\

\multicolumn{1}{l|}{}                      
&     
& \multicolumn{1}{c}{srcml} 
& \multicolumn{1}{c}{0.018} 
& \multicolumn{1}{c}{0.46}  
\\

\multicolumn{1}{l|}{}                      
&    
& \multicolumn{1}{c}{dataflow graph} 
& \multicolumn{1}{c}{0.048} 
& \multicolumn{1}{c}{0.57} 
\\ 

\multicolumn{1}{l|}{}                      
&     
& \multicolumn{1}{c}{callgraph} 
& \multicolumn{1}{c}{0.018} 
& \multicolumn{1}{c}{0.54} 
\\ \hline

\end{tabular}
\vspace{-2mm}
\end{table}

\begin{table}[]
\centering
\small
\caption{BLEU score for code translation}
\label{tab:rq1_codetrans_results}
\begin{tabular}{lllll}
\multicolumn{1}{l|}{} & model & \multicolumn{1}{c}{config} & \multicolumn{1}{c}{BLEU} & \multicolumn{1}{c}{p} \\ \hline

\multicolumn{1}{l|}{\multirow{8}{*}{Java}} & \multirow{4}{*}{jam} & \multicolumn{1}{c}{base} & \multicolumn{1}{c}{50.34} & \multicolumn{1}{c}{-} \\
\multicolumn{1}{l|}{} & & \multicolumn{1}{c}{srcml} & \multicolumn{1}{c}{49.26} & \multicolumn{1}{c}{0.90} \\
\multicolumn{1}{l|}{} & & \multicolumn{1}{c}{dataflow graph} & \multicolumn{1}{c}{50.25} & \multicolumn{1}{c}{0.08} \\
\multicolumn{1}{l|}{} & & \multicolumn{1}{c}{callgraph} & \multicolumn{1}{c}{-} & \multicolumn{1}{c}{-} \\
\cline{2-5}

\multicolumn{1}{l|}{} & \multirow{4}{*}{codellama} & \multicolumn{1}{c}{base} & \multicolumn{1}{c}{80.36} & \multicolumn{1}{c}{-} \\
\multicolumn{1}{l|}{} & & \multicolumn{1}{c}{srcml} & \multicolumn{1}{c}{80.52} & \multicolumn{1}{c}{0.56} \\
\multicolumn{1}{l|}{} & & \multicolumn{1}{c}{dataflow graph} & \multicolumn{1}{c}{79.82} & \multicolumn{1}{c}{0.57} \\
\multicolumn{1}{l|}{} & & \multicolumn{1}{c}{callgraph} & \multicolumn{1}{c}{81.84} & \multicolumn{1}{c}{$p<0.01$} \\ \hline

\multicolumn{1}{l|}{\multirow{4}{*}{C}} & \multirow{4}{*}{codellama} & \multicolumn{1}{c}{base} & \multicolumn{1}{c}{37.82} & \multicolumn{1}{c}{-} \\
\multicolumn{1}{l|}{} & & \multicolumn{1}{c}{srcml} & \multicolumn{1}{c}{37.50} & \multicolumn{1}{c}{0.84} \\
\multicolumn{1}{l|}{} & & \multicolumn{1}{c}{dataflow graph} & \multicolumn{1}{c}{41.64} & \multicolumn{1}{c}{0.91} \\
\multicolumn{1}{l|}{} & & \multicolumn{1}{c}{callgraph} & \multicolumn{1}{c}{39.70} & \multicolumn{1}{c}{0.91} \\ \hline

\end{tabular}
\vspace{-3mm}
\end{table}

\begin{table}[]
\vspace{3mm}
    \centering
    \small
    \caption{Execution-based evaluation on code translation}
    \label{tab:rq1_codetrans_execution_results}
\begin{tabular}{lllllll} 
\multicolumn{1}{l|}{}       & model & \multicolumn{1}{c}{config} & \multicolumn{1}{c}{Pass@1} & \multicolumn{1}{c}{Compilation rate}  \\ \hline
\multicolumn{1}{l|}{\multirow{4}{*}{Java}}                  
& \multirow{4}{*}{codellama}  & \multicolumn{1}{c}{base} & \multicolumn{1}{c}{0.0} & \multicolumn{1}{c}{0.60} \\
\multicolumn{1}{l|}{}                      &     & \multicolumn{1}{c}{srcml} & \multicolumn{1}{c}{0.0} & \multicolumn{1}{c}{0.51}  \\
\multicolumn{1}{l|}{}                      &    & \multicolumn{1}{c}{dataflow graph} & \multicolumn{1}{c}{0.0} & \multicolumn{1}{c}{0.52} \\ 
\multicolumn{1}{l|}{}                      &     & \multicolumn{1}{c}{callgraph} & \multicolumn{1}{c}{0.0} & \multicolumn{1}{c}{0.69} \\\hline

\end{tabular}
\vspace{-3mm}
\end{table}

\begin{table}[b]
\small
    \centering
    \caption{Results for srcml generated by models finetuned with code intelligence tasks.}
    \label{tab:srcml_codeinteligence_result}
\begin{tabular}{llll} &   &     & \multicolumn{1}{c}{Metrics}                  \\
\multicolumn{1}{l|}{}       & task & \multicolumn{1}{c}{model} & \multicolumn{1}{c}{Levenshtein}                         \\ \hline

\multicolumn{1}{l|}{\multirow{3}{*}{Java}}                      & \multirow{3}{*}{srcml}  & \multicolumn{1}{c}{codesum} & \multicolumn{1}{c}{0.350}         \\
\multicolumn{1}{l|}{}                      &     & \multicolumn{1}{c}{codegen}                        & \multicolumn{1}{c}{0.059}      \\
\multicolumn{1}{l|}{}                      &    & \multicolumn{1}{c}{codetrans}                            & \multicolumn{1}{c}{0.430}           \\ 
\hline
\multicolumn{1}{l|}{\multirow{3}{*}{C}}                      & \multirow{3}{*}{srcml}  & \multicolumn{1}{c}{codesum} & \multicolumn{1}{c}{0.035}         \\
\multicolumn{1}{l|}{}                      &     & \multicolumn{1}{c}{codegen}                        & \multicolumn{1}{c}{0.133}      \\
\multicolumn{1}{l|}{}                      &    & \multicolumn{1}{c}{codetrans}                            &   \multicolumn{1}{c}{0.243}        \\

\end{tabular}

\end{table}

\begin{table}[t]
\centering
\small

\caption{Example of error types analysis for callgraph.}
\label{tab:callgraphqulatative}

\begin{tabular}{ll}
\end{tabular}

\begin{tabular}{c}
\begin{lstlisting}[language=Java, basicstyle=\ttfamily\small, breaklines=true, showstringspaces=false]

public void setAlignment(int row, int column,
    HorizontalAlignmentConstant hAlign,  VerticalAlignmentConstant vAlign){
    prepareCell(row, column);
    setHorizontalAlignment(row, column, hAlign);
    setVerticalAlignment(row, column, vAlign);
}

public void setHorizontalAlignment(int row, int column, HorizontalAlignmentConstant align){
    prepareCell(row, column);
    DOM.setStyleAttribute(getCellElement(tableElem, row,  column),
        "textAlign", align.getTextAlignString());
}

protected void prepareCell(int row, int column){
    prepareRow(row);
    if (column < 0){
      throw new IndexOutOfBoundsException(
        "Cannot create a column with a negative index: " + column);
    }

    // Ensure that the requested column exists.
    int cellCount = getCellCount(row);
    for (int i = cellCount; i <= column; ++i)
      addCell(row);
}

\end{lstlisting} \\

\end{tabular}

\begin{tabular}{l|p{10cm}}
\hline
reference &\begin{lstlisting}[language=C, basicstyle=\ttfamily\scriptsize, breaklines=true, showstringspaces=false]
1. setAlignment -> prepareCell
2. setAlignment -> setHorizontalAlignment -> prepareCell

\end{lstlisting} \\\hline
     gpt &  \begin{lstlisting}[language=Java, basicstyle=\ttfamily\scriptsize, breaklines=true, showstringspaces=false]
1. setAlignment->prepareCell->prepareRow
2. setAlignment->prepareCell->getCellCount
3. setAlignment->prepareCell->addCell
4. setAlignment->setHorizontalAlignment
5. setAlignment->setVerticalAlignment


\end{lstlisting}\\\hline

\end{tabular}
\end{table}

\begin{table*}[h]
\centering
\rotatebox{90}{%
\small
    \begin{tabular}{llllll}
    &   &     & \multicolumn{3}{c}{Metrics} \\
\multicolumn{1}{l|}{}       & task & \multicolumn{1}{c}{model} & \multicolumn{1}{c}{Jaccard similarity}      &  \multicolumn{1}{c}{pair accuracy}     &  \multicolumn{1}{c}{chain accuracy} \\ \hline

\multicolumn{1}{l|}{\multirow{6}{*}{Java}} & \multirow{3}{*}{callgraph}  & \multicolumn{1}{c}{codesum} & \multicolumn{1}{c}{0.061} & \multicolumn{1}{c}{0.076}  & \multicolumn{1}{c}{0} \\
\multicolumn{1}{l|}{}                      &     & \multicolumn{1}{c}{codegen} & \multicolumn{1}{c}{ 0.011} & \multicolumn{1}{c}{0.046} & \multicolumn{1}{c}{1.07E-5} \\
\multicolumn{1}{l|}{}                      &    & \multicolumn{1}{c}{codetrans} & \multicolumn{1}{c}{0.013} & \multicolumn{1}{c}{0.050}   & \multicolumn{1}{c}{0} \\ 

 \cline{2-6}

\multicolumn{1}{l|}{}                      & \multirow{3}{*}{dataflow graph}  & \multicolumn{1}{c}{codesum} & \multicolumn{1}{c}{0.0019} & \multicolumn{1}{c}{0.0038}  & \multicolumn{1}{c}{4.96E-04} \\
\multicolumn{1}{l|}{}                      &     & \multicolumn{1}{c}{codegen} & \multicolumn{1}{c}{0.0} & \multicolumn{1}{c}{0.0} & \multicolumn{1}{c}{0.0} \\
\multicolumn{1}{l|}{}                      &    & \multicolumn{1}{c}{codetrans} & \multicolumn{1}{c}{0.0044} & \multicolumn{1}{c}{0.0072}   & \multicolumn{1}{c}{1.44E-03} \\ 


\hline

\multicolumn{1}{l|}{\multirow{6}{*}{C}} & \multirow{3}{*}{callgraph}  & \multicolumn{1}{c}{codesum} & \multicolumn{1}{c}{0.002} & \multicolumn{1}{c}{0.025}  & \multicolumn{1}{c}{0.00012} \\
\multicolumn{1}{l|}{}                      &     & \multicolumn{1}{c}{codegen} & \multicolumn{1}{c}{0.048} & \multicolumn{1}{c}{0.075} & \multicolumn{1}{c}{2.64E-06} \\
\multicolumn{1}{l|}{}                      &    & \multicolumn{1}{c}{codetrans} & \multicolumn{1}{c}{0.016} & \multicolumn{1}{c}{0.028}   & \multicolumn{1}{c}{2.14E-07} \\ 

\cline{2-6}

\multicolumn{1}{l|}{}                      & \multirow{3}{*}{dataflow graph}  & \multicolumn{1}{c}{codesum} & \multicolumn{1}{c}{0.0} & \multicolumn{1}{c}{0.0}  & \multicolumn{1}{c}{0.0} \\
\multicolumn{1}{l|}{}                      &     & \multicolumn{1}{c}{codegen} & \multicolumn{1}{c}{0.0} & \multicolumn{1}{c}{0.0} & \multicolumn{1}{c}{0.0} \\
\multicolumn{1}{l|}{}                      &    & \multicolumn{1}{c}{codetrans} & \multicolumn{1}{c}{1.36E-06} & \multicolumn{1}{c}{4.91E-06}   & \multicolumn{1}{c}{0.0} \\ 


    \end{tabular}
}
\caption{Results for callgraph and dataflow graph generated by the models finetuned with various code intelligence tasks.}
\label{tab:callgraph_datagraph_code_intelligence}
\end{table*}

\subsection{RQ4: Open Models, Dev. + Static Analysis}
We observed significantly lower results compared with models finetuned to do the tasks. We showed the results for AST generation in Table~\ref{tab:srcml_codeinteligence_result} and callgraph and dataflow graph generation in Table~\ref{tab:callgraph_datagraph_code_intelligence}. This shows that LLMs do not use static analysis tasks for code intelligence tasks. For example, in Table~\ref{tab:codesumcallgraphexample}, we observed that ``setHorizontalAlignment'' does not call ``setVerticalAlignment'', but CodeLlaMA generates a path where ``setHorizontalAlignment'' is calling ``setVerticalAlignment''.  This shows that LLMs do not have the fundamental understanding of callgraph generation even though it has already learnt to generate code summarization. Another possible explanation is that LLMs could learn the form from the downstream tasks instead of the semantic meaning. For example,~\cite{bender2020climbing} showed that LLMs could learn both the ``form'' of the language and the ``semantics'' of the language. They observed that LLMs do not need to learn the semantics of the language to have good performance on code development tasks. LLMs can do the downstream tasks well while learning the form from the training data only. This result aligns with the finding that LLMs only memorize the form in the training data instead of understanding the semantics of the training data~\citep{ma2025memorization}. This would guide future research to develop new model architectures or new approaches to train models. 

\begin{table}[t]
\centering
\small

\caption{Example that shows that LLMs still struggle with generating correct call path after we finetuned it with code summarization.}
\label{tab:codesumcallgraphexample}
\vspace{-5mm}
\begin{tabular}{ll}
\end{tabular}

\begin{tabular}{c}
\begin{lstlisting}[language=Java, basicstyle=\ttfamily\small, breaklines=true, showstringspaces=false]

public void setAlignment(int row, int column,
    HorizontalAlignmentConstant hAlign,  VerticalAlignmentConstant vAlign){
    prepareCell(row, column);
    setHorizontalAlignment(row, column, hAlign);
    setVerticalAlignment(row, column, vAlign);
}

public void setHorizontalAlignment(int row, int column, HorizontalAlignmentConstant align){
    prepareCell(row, column);
    DOM.setStyleAttribute(getCellElement(tableElem, row,  column),
        "textAlign", align.getTextAlignString());
}

protected void prepareCell(int row, int column){
    prepareRow(row);
    if (column < 0){
      throw new IndexOutOfBoundsException(
        "Cannot create a column with a negative index: " + column);
    }

    // Ensure that the requested column exists.
    int cellCount = getCellCount(row);
    for (int i = cellCount; i <= column; ++i)
      addCell(row);
}

\end{lstlisting} \\

\end{tabular}

\begin{tabular}{l|p{10cm}}
\hline
reference &\begin{lstlisting}[language=C, basicstyle=\ttfamily\scriptsize, breaklines=true, showstringspaces=false]
1. setAlignment -> prepareCell
2. setAlignment -> setHorizontalAlignment -> prepareCell

\end{lstlisting} \\\hline
     codellama &  \begin{lstlisting}[language=Java, basicstyle=\ttfamily\scriptsize, breaklines=true, showstringspaces=false]
1. setAlignment -> prepareCell -> setHorizontalAlignment -> setVerticalAlignment -> setAttr


\end{lstlisting}\\\hline

\end{tabular}
\end{table}

\newpage

\subsection{RQ5: Analysis of Error Types}
We found that extra direct calls is the common error type followed by missing direct calls and cycles in Java dataset. In C dataset, we found that missing direct calls is the common error types followed by extra direct calls and missing indirect calls. We show the results for callgraph in Table~\ref{tab:callgraph_qualitative_result}. In the Java dataset, we observed that LLMs tend to predict the extra direct calls in callgraph. The possible explanation is that direct calls are more straightforward than indirect edges. This phenomenon can also be verified in the number of functions with cycles. For example, in Java, we only observed three functions with cycles in the reference callgraph. However, we observed 1,510 functions with cycles in the predicted callgraph in~\texttt{gpt-base}. We observed that finetuning CodeLlaMA improves  overall results and reduces the number of cycles. However, we do not observe the reasoning prompt helps to improve the results. Regarding C/C++ dataset, we observed that callgraph generated by LLMs tends to miss more direct edges and finetuning CodeLlaMA improves the results. Note that it is possible that one function has extra direct edges and misses direct edges based on our calculation. We show an example in Table~\ref{tab:callgraphqulatative}. When we compared path 1 in the reference and path 1 in the GPT predicted path, we have extra direct edges, i.e. from prepareCell to prepareRow, whereas we have one missing direct edge  when we compared path 1 in the reference and path 4 in the GPT predicted path.

\begin{table*}[h!]
\centering
\rotatebox{90}{%
\small

\begin{tabular}{llllllll}
 & & & \multicolumn{5}{c}{Metrics} \\
\multicolumn{1}{l|}{} & task & model & MDC & MIC & RC & PC & EDC \\ \hline

\multicolumn{1}{l|}{\multirow{7}{*}{Java}} & \multirow{7}{*}{callgraph} & gpt-base & 10,246 & 68 & 3 & 1,510 & 11,735 \\
\multicolumn{1}{l|}{} & {} & gpt-in-context & 10,316 & 68 & 3 & 1,243 & 11,985 \\
\multicolumn{1}{l|}{} & {} & gpt-in-context-reason & 10,480 & 65 & 3 & 1,297 & 12,120 \\ 
\multicolumn{1}{l|}{} & {} & gemini-base & 10,480 & 65 & 3 & 1,297 & 12,120 \\ 
\multicolumn{1}{l|}{} & {} & gemini-in-context & 9,855 & 65 & 3 & 1,087 & 10,704 \\ 
\multicolumn{1}{l|}{} & {} & gemini-in-context-reason & 10,146 & 64 & 3 & 534 & 9,953 \\ 
\multicolumn{1}{l|}{} & {} & codellama-finetuned & 3,256 & 65 & 3 & 43 & 2,738 \\ \cline{1-8}

\multicolumn{1}{l|}{\multirow{7}{*}{C}} & \multirow{7}{*}{callgraph} & gpt-base & 21,773 & 10,863 & 82 & 4,075 & 20,016 \\
\multicolumn{1}{l|}{} & {} & gpt-in-context & 22,662 & 11,170 & 82 & 1,598 & 20,886 \\
\multicolumn{1}{l|}{} & {} & gpt-in-context-reason & 22,953 & 11,200 & 82 & 1,722 & 21,480 \\
\multicolumn{1}{l|}{} & {} & gemini-base & 22,589 & 11,049 & 82 & 1,130 & 20,430 \\ 
\multicolumn{1}{l|}{} & {} & gemini-in-context & 21,551 & 10,899 & 82 & 1,033 & 19,203 \\ 
\multicolumn{1}{l|}{} & {} & gemini-in-context-reason & 21,883 & 11,241 & 82 & 476 & 14,422 \\ 
\multicolumn{1}{l|}{} & {} & codellama-finetuned & 18,928 & 11,169 & 82 & 649 & 10,591
\end{tabular}
}
\caption{Error types analysis for callgraph generation; MDC stands for missing direct call; EDC stands for extra direct call; MIC stands for missing indirect call; RC stands for reference cycles; PC stands for predicted cycles.}
\label{tab:callgraph_qualitative_result}
\end{table*}

For dataflow analysis in Java, we found that extra direct edges is the most common error followed by extra indirect edges. In C dataset, we observed that extra direct edges is the most common error followed by missing direct edges. We showed the results for dataflow analysis in Table~\ref{tab:dataflow_qualitative_result}. The possible explanation is that LLMs are lack of semantic reasoning skills to generate the correct edges. We showed one example for C in Table~\ref{tab:dataflowqulatative}. We found that the reference only has one path, whereas LLMs generate multiple paths and all of the paths are incorrect. This example would be counted in both extra direct edges and missing edges. This result aligns with the findings by~\cite{ma2023lms,xie2025core} that LLMs tend to hallucinate when doing more complex reasoning tasks.

We observed that tag mismatch is the most common error in LLMs generated AST and in-context learning reduces the number of functions with tag mismatches. We show the results for AST in Table~\ref{tab:srcml_qualitative_result}. For example, we observed 7,135 functions have tag mismatches in~\texttt{gpt-base}, whereas we only observed 703 functions that have tag mismatches in~\texttt{gpt-in-context}. The possible explanations are 1) LLMs are lack of semantic reasoning skills, so LLMs cannot generate the proper tags based on the semantic information. and 2) LLMs do not know the proper SrcML format, so in-context learning helps LLMs to learn the proper format. For example, we observed that the tools use~\texttt{decl} as the tag whereas LLMs use~\texttt{expr} as the tag. To our surprise,  we found that finetuning improves the Levenshtein, but we observe the lower compile success rate. The explanation is that we limit the output length of CodeLlama to 512 tokens due to computational cost. We showed one example in Table~\ref{tab:srcmlcodellamaqualitativeexample} that the SrcML is incomplete. Overall, our results align with our previous findings that LLMs are lack of semantic reasoning skills. 

\subsection{Discussion}

The key novelty of this paper is that we explore whether LLMs use static analysis to do code development tasks like a human programmer would. Recent research has shown that pretraining LLMs with large source code and increasing model size improve code development tasks~\citep{wang2021codet5,roziere2023code, li2023starcoder}. However, research has shown that programmers build mental models of program behavior through static analysis~\citep{Jiang2016,beller2016analyzing, vassallo2020developers, wallace2025programmer}. One would expect that pretraining LLMs with static analysis tasks would also improve the code development tasks if LLMs use the same thought process as human programmers. One would also expect LLMs have static analysis as a byproduct if LLMs use static analysis while doing code development tasks. Therefore, we conducted exhaustive experiments  to answer this question. Although several papers have more thorough discussion and define more thorough taxonomies on the error types of the tasks~\citep{sun2024source, pan2024lost,yuan2023evaluating, jiang2024survey, ma2023lms}, those papers mainly focus on the discussion of individual tasks without building the connection between these two tasks. Most of the current papers that discuss the connection between these two tasks only focus on how to include static analysis  in the models~\citep{alon2018code2seq, sun2020treegen, leclair2020improved, sun2020treegen, tang2022ast, bansal2023function, du2023pre} and in the prompts~\citep{wang2024llmdfa, su2025context, luo2025can, zhang2026systematic} to improve code development tasks. Providing static analysis in the input would give models opportunities to learn the ``form'' of the programming languages instead of understanding whether models use the ``semantics'' of static analysis.  Therefore, we explore this question in this paper.


We found that  LLMs do not do static analysis. Specifically, we have two key observations. First, training models with static analysis tasks does not improve the code development tasks and training models with code development tasks does not lead to models being able to perform static analysis tasks. This would imply that LLMs do not use static analysis tasks while doing code development tasks and LLMs do not have an  understanding of the semantics of static analysis. Second, LLMs could make the same logical mistakes as in the models without being pretrained with static analysis tasks. This shows  that LLMs learn the ``form''  instead of ``semantics''. This finding suggests future research developing new model architectures and training procedures that mimic what human programmers would do  for software engineering tasks.



\begin{table*}[h!]
\centering
\rotatebox{90}{%
\scriptsize
    \begin{tabular}{lllllll}
    &   &     & \multicolumn{4}{c}{Metrics} \\
\multicolumn{1}{l|}{}       & task & \multicolumn{1}{c}{model} & \multicolumn{1}{c}{EDE}      &  \multicolumn{1}{c}{MDE}     &  \multicolumn{1}{c}{MIE} &  \multicolumn{1}{c}{EIE} \\ \hline

\multicolumn{1}{l|}{\multirow{7}{*}{Java}} & \multirow{7}{*}{dataflow}  & \multicolumn{1}{c}{gpt-base} & \multicolumn{1}{c}{7,042} & \multicolumn{1}{c}{7,041}  & \multicolumn{1}{c}{5,602}& \multicolumn{1}{c}{7,042}\\
\multicolumn{1}{l|}{}                      &     & \multicolumn{1}{c}{gpt-in-context} & \multicolumn{1}{c}{6,997} & \multicolumn{1}{c}{6,436} & \multicolumn{1}{c}{4,501}& \multicolumn{1}{c}{6,934}\\
\multicolumn{1}{l|}{}                      &     & \multicolumn{1}{c}{gpt-in-context-reason} & \multicolumn{1}{c}{7,002} & \multicolumn{1}{c}{6,487} & \multicolumn{1}{c}{4,502}& \multicolumn{1}{c}{6,940}\\
\multicolumn{1}{l|}{}                      &    & \multicolumn{1}{c}{gemini-base} & \multicolumn{1}{c}{7,036} & \multicolumn{1}{c}{7,033}   & \multicolumn{1}{c}{5,602}& \multicolumn{1}{c}{7,035} \\ 
\multicolumn{1}{l|}{}                      &     & \multicolumn{1}{c}{gemini-in-context} & \multicolumn{1}{c}{6,809} & \multicolumn{1}{c}{6,399} & \multicolumn{1}{c}{4,711}& \multicolumn{1}{c}{6,554}\\ 
\multicolumn{1}{l|}{}                      &     & \multicolumn{1}{c}{gemini-in-context-reason} & \multicolumn{1}{c}{6,688} & \multicolumn{1}{c}{6,284} & \multicolumn{1}{c}{4,578}& \multicolumn{1}{c}{6,200}\\
\multicolumn{1}{l|}{}                      &     & \multicolumn{1}{c}{codellama-finetuned} & \multicolumn{1}{c}{6,119} & \multicolumn{1}{c}{5,442}    & \multicolumn{1}{c}{4,756}  & \multicolumn{1}{c}{6,087} \\\cline{1-7}

\multicolumn{1}{l|}{\multirow{7}{*}{C}} & \multirow{7}{*}{dataflow}  & \multicolumn{1}{c}{gpt-base} & \multicolumn{1}{c}{8,128} & \multicolumn{1}{c}{8,104}  & \multicolumn{1}{c}{5,439} & \multicolumn{1}{c}{8,128}\\
\multicolumn{1}{l|}{}                      &     & \multicolumn{1}{c}{gpt-in-context} & \multicolumn{1}{c}{7,957} & \multicolumn{1}{c}{8,042} & \multicolumn{1}{c}{5,456} & \multicolumn{1}{c}{7,922}  \\
\multicolumn{1}{l|}{}                      &     & \multicolumn{1}{c}{gpt-in-context-reason} & \multicolumn{1}{c}{7,964} & \multicolumn{1}{c}{8,053} & \multicolumn{1}{c}{5,458}& \multicolumn{1}{c}{7,951}\\
\multicolumn{1}{l|}{}                      &    & \multicolumn{1}{c}{gemini-base} & \multicolumn{1}{c}{8,093} & \multicolumn{1}{c}{8,084}   & \multicolumn{1}{c}{5,445} & \multicolumn{1}{c}{8,092}\\ 
\multicolumn{1}{l|}{}                      &     & \multicolumn{1}{c}{gemini-in-context} & \multicolumn{1}{c}{7,763} & \multicolumn{1}{c}{7,960} & \multicolumn{1}{c}{5,397}& \multicolumn{1}{c}{7,700}\\
\multicolumn{1}{l|}{}                      &     & \multicolumn{1}{c}{gemini-in-context-reason} & \multicolumn{1}{c}{7,962} & \multicolumn{1}{c}{7,978} & \multicolumn{1}{c}{5,403}& \multicolumn{1}{c}{7,902}\\
\multicolumn{1}{l|}{}                      &     & \multicolumn{1}{c}{codellama-finetuned} & \multicolumn{1}{c}{5,770} & \multicolumn{1}{c}{6,497}    & \multicolumn{1}{c}{5,046} & \multicolumn{1}{c}{5,578}

    \end{tabular}
}
\caption{Error types analysis for dataflow generation; EDE stands for extra direct edges; MDE stands for missing direct edges; MIE stands for missing indirect edges; EIE stands for extra indirect edges}
\label{tab:dataflow_qualitative_result}
\end{table*}

\begin{table*}[h]
\centering
\rotatebox{90}{%
\scriptsize
    \begin{tabular}{lllllll}
    &   &     & \multicolumn{4}{c}{Metrics} \\
\multicolumn{1}{l|}{}       & task & \multicolumn{1}{c}{model} & \multicolumn{1}{c}{MN}      &  \multicolumn{1}{c}{EN}     &  \multicolumn{1}{c}{TM}   &  \multicolumn{1}{c}{PE}\\ \hline

\multicolumn{1}{l|}{\multirow{5}{*}{Java}} & \multirow{5}{*}{srcml} & \multicolumn{1}{c}{gpt-base} & \multicolumn{1}{c}{5,380} & \multicolumn{1}{c}{2,428}  & \multicolumn{1}{c}{7,135}& \multicolumn{1}{c}{1,498}\\
\multicolumn{1}{l|}{}                      &     & \multicolumn{1}{c}{gpt-in-context} & \multicolumn{1}{c}{971} & \multicolumn{1}{c}{1,036} & \multicolumn{1}{c}{703}  & \multicolumn{1}{c}{7,249}\\
\multicolumn{1}{l|}{}                      &    & \multicolumn{1}{c}{gemini-base} & \multicolumn{1}{c}{0} & \multicolumn{1}{c}{0}   & \multicolumn{1}{c}{0} & \multicolumn{1}{c}{8,625}\\ 
\multicolumn{1}{l|}{}                      &     & \multicolumn{1}{c}{gemini-in-context} & \multicolumn{1}{c}{1,412} & \multicolumn{1}{c}{4,253} & \multicolumn{1}{c}{1,541}& \multicolumn{1}{c}{3,876}\\
\multicolumn{1}{l|}{}                      &     & \multicolumn{1}{c}{codellama-finetuned} & \multicolumn{1}{c}{21} & \multicolumn{1}{c}{10}    & \multicolumn{1}{c}{5}& \multicolumn{1}{c}{8,282}\\\cline{1-7}

\multicolumn{1}{l|}{\multirow{5}{*}{C}} &  \multirow{5}{*}{srcml} & \multicolumn{1}{c}{gpt-base} & \multicolumn{1}{c}{4,216}   & \multicolumn{1}{c}{3,398}& \multicolumn{1}{c}{5,988}& \multicolumn{1}{c}{3,567}\\
\multicolumn{1}{l|}{}                      &     & \multicolumn{1}{c}{gpt-in-context} & \multicolumn{1}{c}{7} & \multicolumn{1}{c}{11} & \multicolumn{1}{c}{4}& \multicolumn{1}{c}{9,549}\\
\multicolumn{1}{l|}{}                      &    & \multicolumn{1}{c}{gemini-base} & \multicolumn{1}{c}{2,719} & \multicolumn{1}{c}{5,824}  & \multicolumn{1}{c}{6,313}& \multicolumn{1}{c}{2,690} \\ 
\multicolumn{1}{l|}{}                      &     & \multicolumn{1}{c}{gemini-in-context} & \multicolumn{1}{c}{2,437} & \multicolumn{1}{c}{5,432} & \multicolumn{1}{c}{3,941}& \multicolumn{1}{c}{3,285}\\ 

\multicolumn{1}{l|}{}                      &     & \multicolumn{1}{c}{codellama-finetuned} & \multicolumn{1}{c}{2}    & \multicolumn{1}{c}{3} & \multicolumn{1}{c}{2}  & \multicolumn{1}{c}{9,549}

    \end{tabular}
}
\caption{Error types analysis for AST generation; MN stands for missing nodes; EN stands for extra nodes; TM stands for tag mismatches; PE stands for parsing errors.}
\label{tab:srcml_qualitative_result}
\end{table*}

\begin{table}[!h]
\centering
\scriptsize
\vspace{-3mm}
\caption{Example of error types analysis for AST generation in CodeLlama.}
\label{tab:srcmlcodellamaqualitativeexample}
\vspace{-9mm}
\begin{tabular}{ll}
\end{tabular}

\begin{tabular}{p{\dimexpr\linewidth-2\tabcolsep}}
\begin{lstlisting}[language=Java, basicstyle=\ttfamily\tiny, breaklines=true, showstringspaces=false, xleftmargin=0pt, breakindent=0pt]

<?xml version="1.0" encoding="UTF-8" standalone="yes"?><unit xmlns="http://www.srcML.org/srcML/src" evision="1.0.0" language="Java" filename="temp.java"> <function><type><specifier>private</specifier> <name>TreeItem</name></type> <name>findNodeWithIdRecursive</name><parameter_list>(<parameter><decl><type><name><name>TreeItem</name><index>[]</index></name></type> <name>arrItem</name></decl></parameter>, <parameter><decl><type><name>String</name></type> <name>id</name></decl></parameter>)</parameter_list><block>{<block_content><for>for<control>(<init><decl><type><name>TreeItem</name></type> <name>current</name><range>: <expr><name><name>arrItem</name><index>[]</index></name></expr></range></decl></init>)</control><block>{<block_content><decl_stmt><decl><type><name>String</name></type> <name>idInNode</name> <init>= <expr><operator>(</operator><name>String</name><operator>)</operator> <call><name><name>current</name><operator>.</operator><name>getData</name></name><argument_list>()</argument_list></call></expr></init></decl>;</decl_stmt><if_stmt><if>if<condition>(<expr><call><name><name>idInNode</name><operator>.</operator><name>equals</name></name><argument_list>(<argument><expr><name>id</name></expr></argument>)</argument_list></call></expr>)</condition><block type="pseudo"><block_content><return>return <expr><name>current</name></expr>;</return></block_content></block></if></if_stmt><decl_stmt><decl><type><name>TreeItem</name></type> <name>itemInside</name> <init>= <

\end{lstlisting} \\

\end{tabular}

\end{table}

\vspace{-3mm}
\section{Conclusion}

In this paper, we have four key observations. First, we found that in-context learning and finetuning only improve the performance on AST generation. However, in-context learning does not improve more complex static analysis tasks and finetuning  only improves the more complex static analysis tasks partially. Specifically, we do not observe any significant improvement on callgraph generation and dataflow graph generation with in-context learning. Moreover, we found that finetuning improves pair accuracy and Jaccard similarity although we do not observe a significant improvement on chain accuracy. This result shows that LLMs struggle with more complex reasoning.

Second, we found that pretraining models with various static analysis tasks do not necessarily improve the code development tasks. Specifically, we do not strongly observe the statistical difference across all datasets and all programming languages when we pretrain models with different static analysis tasks. Another piece of evidence that shows that static analysis tasks do not improve code development tasks is that LLMs tend to make the same logical errors after pretraining models with static analysis tasks. This shows that LLMs do not use static analysis as human programmers would.

Third, we found that LLMs do not have static analysis as a byproduct when we train LLMs with code development tasks. We observed significantly lower results compared with models specifically finetuned or prompted to do the tasks. The possible explanation is that LLMs learn the form of the training data instead of the semantics of the training data. In other words, LLMs could simply memorize the input form instead of understanding the semantic information of the input.

Finally, we observed that LLMs tend to generate outputs with simple semantic relations. For example, we found that LLMs tend to predict more cycles than it actually has in the callgraph. The possible explanation is that cycles are more straightforward for LLMs to predict than predicting indirect edges. This would further show that LLMs are lack of semantic reasoning skills and guide future research to develop new model architectures and training procedures to help LLMs to learn the semantics of inputs. Overall, we found that LLMs do not use static analysis like a human would.

\begin{acknowledgements}
This work is supported in part by the NSF grants CCF-2100035, CCF-2211428, and CCF-2211429. 
Any opinions, findings, and conclusions expressed herein are the authors’ and do not necessarily reflect those of the sponsors.
\end{acknowledgements}

\section*{Ethical approval} 
This study was conducted in accordance with institutional guidelines and approved by the Institutional Review Board.

\section*{Informed consent}
Not applicable.

\section*{Author Contributions}  
Chia-Yi Su designed the experiments, implemented the code for the experiments, and wrote the manuscript. Collin McMillan wrote the manuscript and designed the experiments.

\section*{Code / Data Availability Statement}
We release all data, scripts, survey materials, and results at {\url{https://github.com/apcl-research/llm-reason}}.

\section*{Clinical trial number}
Not applicable

%
\section*{Conflict of interest}
The authors declare that they have no conflict of interest.

\bibliographystyle{spbasic}      
\bibliography{main}

\end{document}